\documentclass[a4paper,11pt]{article}

\usepackage{jcappub} 

\usepackage[utf8]{inputenc}  
\usepackage[T1]{fontenc}     
\usepackage{amsmath,amssymb,lmodern} 
\usepackage{sidecap}
\usepackage[caption=false]{subfig}
\usepackage{graphicx}
\usepackage{tabularx}
\usepackage{enumerate} 
\usepackage{makecell}
\usepackage{hyperref}\hypersetup{
    colorlinks,%
    citecolor=blue,%
    filecolor=blue,%
    linkcolor=blue,%
    urlcolor=blue
}
\usepackage{color}
\usepackage{cleveref}
\allowdisplaybreaks[1]

\newcommand{\newc}{\newcommand}
\newc{\tif}{\tilde{f}}
\newc{\tih}{\tilde{h}}
\newc{\tip}{\tilde{\phi}}
\newc{\tiA}{\tilde{A}}

\newcommand{\mr}[1]{\mathrm{#1}}
\newcommand{\mL}[1]{\mathcal{#1}}

\newcommand{\ben}{\begin{eqnarray}}
\newcommand{\een}{\end{eqnarray}}
\newc{\be}{\begin{equation}}
\newc{\ee}{\end{equation}}
\newc{\ba}{\begin{eqnarray}}
\newc{\ea}{\end{eqnarray}}
\newc{\bea}{\begin{eqnarray*}}
\newc{\eea}{\end{eqnarray*}}
\newc{\D}{\partial}
\newc{\ie}{{\it i.e.} }
\newc{\eg}{{\it e.g.} }
\newc{\etc}{{\it etc.} }
\newc{\etal}{{\it et al.}}

\newc{\ra}{\rightarrow}
\newc{\lra}{\leftrightarrow}
\newc{\lsim}{\buildrel{<}\over{\sim}}
\newc{\gsim}{\buildrel{>}\over{\sim}}
\newc{\aP}{\alpha_{\rm P}}
\newc{\dphi}{\delta\phi}
\newc{\da}{\delta A}
\newc{\tp}{\dot{\phi}}
\newc{\Ve}{V_{\rm eff}}
\newc{\Vep}{V_{{\rm eff},\phi}}

\title{\boldmath Anisotropic inflation with coupled $p-$forms}

\author[a]{Juan P. Beltr\'an Almeida,}
\author[a,b]{Alejandro Guarnizo,}
\author[c]{Ryotaro Kase,}
\author[c]{Shinji Tsujikawa,}
\author[b]{and C\'esar A. Valenzuela-Toledo}

\affiliation[a]{Departamento de F\'isica, Facultad de Ciencias, Universidad Antonio Nari\~no, 
Cra 3 Este \# 47A-15, Bogot\'a D.C. 110231, Colombia}
\affiliation[b]{Departamento de F\'isica, Universidad del Valle, Ciudad Universitaria Mel\'endez, 
Santiago de Cali 760032, Colombia}
\affiliation[c]{Department of Physics, Faculty of Science, Tokyo University of Science, 1-3, Kagurazaka,
Shinjuku-ku, Tokyo 162-8601, Japan}

\emailAdd{juanpbeltran@uan.edu.co}
\emailAdd{alejandro.guarnizo@correounivalle.edu.co}
\emailAdd{r.kase@rs.tus.ac.jp}
\emailAdd{shinji@rs.kagu.tus.ac.jp}
\emailAdd{cesar.valenzuela@correounivalle.edu.co}

\abstract{We study the cosmology in the presence of arbitrary 
couplings between $p$-forms in 4-dimensional space-time 
for a general action respecting gauge symmetry 
and parity invariance.  
The interaction between 0-form (scalar field $\phi$) and 3-form fields gives 
rise to an effective potential $V_{\rm eff}(\phi)$ for the former 
after integrating out the contribution of the latter. 
We explore the dynamics of inflation on an anisotropic cosmological 
background for a coupled system of 0-, 1-, and 2-forms. 
In the absence of interactions between 1- and 2-forms, we derive
conditions under which the anisotropic shear endowed with nearly 
constant energy densities of 1- and 2-forms 
survives during slow-roll inflation for an arbitrary scalar 
potential $V_{\rm eff}(\phi)$. 
If 1- and 2-forms are coupled to each other, we show the existence 
of a new class of anisotropic inflationary solutions in which 
the energy density of 2-form is sustained by that of 1-form through 
their interactions. Our general analytic formulas for the 
anisotropic shear are also confirmed by the numerical analysis 
for a concrete inflaton potential.
}

\begin{document}
\maketitle
\flushbottom


\section{Introduction}

The observational evidence for inflation and dark energy suggests 
that there may be additional degrees of freedom (DOFs) 
beyond those appearing in standard models of 
particle physics and general relativity. 
The simplest candidate for such new DOFs is a scalar field 
$\phi$ (0-form), which can be compatible with the 
homogeneous and isotropic cosmological background. 
Indeed, the scalar field slowly evolving along a nearly flat potential 
$V(\phi)$ drives the cosmic acceleration \cite{oldinf1,oldinf2,oldinf3,oldinf4,oldinf5,oldinf6,oldinf7}. 
The mechanism of generating scalar and tensor perturbations 
during single-field inflation \cite{oldper1,oldper2,oldper3,oldper4,oldper5,oldper6} is overall consistent with 
the observed scalar spectral index and the tensor-to-scalar 
ratio of Cosmic Microwave Background (CMB)
temperature anisotropies \cite{WMAP,Planck2015,Planck2018}.

On the other hand, it is known that there are some anomalies 
in the CMB data such as the hemispherical asymmetry between  the North 
and South ecliptic hemispheres, the mutual alignment of lowest 
multipole moments, and the dipole modulation of very large-scale 
CMB signals \cite{Planck2018,Planckiso}. 
This may be attributed to the violation of the isotropic 
cosmological evolution. 
For a scalar field, the breaking of isotropy of space-time is 
limited apart from specific cases \cite{Tahara:2018orv}. 
The vector field $A_{\mu}$ (i.e., 1-form field) 
or the 2-form field $B_{\mu \nu}$ can be the natural 
sources for 
generating anisotropies relevant to CMB anomalies.

If we apply the 1-form field to inflation, there is a no-hair 
theorem stating that, in the presence of an effective cosmological constant, 
the anisotropic shear generated by the 1-form dilutes exponentially fast, so, 
an isotropic de Sitter background is quickly obtained \cite{Wald,Starobinsky1982mr}.
Nevertheless, in the presence of the coupling 
$-f_1(\phi) F_{\mu \nu}F^{\mu \nu}/4$, 
where $f_1(\phi)$ is a function of the scalar field $\phi$ and 
$F_{\mu \nu}=\partial_{\mu}A_{\nu}-\partial_{\nu}A_{\mu}$ is 
the field strength of the 1-form $A_{\mu}$, it is possible to 
sustain the anisotropic hair on the quasi de Sitter background 
where the Hubble expansion rate slowly varies in time \cite{Watanabe}. 
The power spectrum of curvature perturbations is modified by 
the broken rotational invariance, whose effect can be quantified 
by an anisotropic parameter $g_*$ \cite{Watanabe2, Gum}. 
The non-linear estimator $f_{\rm NL}$ of scalar non-Gaussianities 
can be as large as the order of 10 in the squeezed limit \cite{Bartolo}. 
Several aspects of the coupling $-f_1(\phi) F_{\mu \nu}F^{\mu \nu}/4$ have been    
discussed in the literature, see Refs.~\cite{Ratra1991bn,Bamba2003av,Martin2007ue,Bamba2008my,Yokoyama2008xw,Watanabe,Watanabe2,Dimopoulos2009vu,Dimopoulos2009am, Himmetoglu2009mk, Gumrukcuoglu2010yc,Namba2012gg,Bartolo,Shiraishi2013vja,Fujita2013pgp,Abolhasani2013zya,Lyth2013kah, Biagetti2013qqa, Chen2014eua,Fujita2017lfu,Fujita:2018zbr}.

The interaction between the scalar $\phi$ and the 2-form field strength 
$H_{\mu \nu \lambda}$ of the form 
$-f_2(\phi)H_{\mu \nu \lambda}H^{\mu \nu \lambda}/12$ can also 
lead to anisotropic inflation for a suitable choice of 
the coupling $f_2(\phi)$. 
In Ref.~\cite{Ohashi1}, it was shown that the power spectrum of curvature
perturbations is of the prolate-type anisotropy with $g_*>0$, in contrast to 
the oblate-type anisotropy for the scalar-vector coupling. 
Moreover the non-linear estimator $f_{\rm NL}$ vanishes in the squeezed limit, 
while it can be of order 10 in the equilateral and enfolded limits as consistent 
with the recent CMB data \cite{Planck2018}. 
Hence the couplings of 1-form 
and 2-form fields with the scalar field $\phi$ can be observationally distinguished from each 
other \cite{Ohashi2} (see also Ref.~\cite{Obata:2018ilf}).

The 3-form field is also relevant to the dynamics of cosmic 
acceleration \cite{Duncan:1989ug,Bousso:2000xa,Dvali:2005an,Kaloper:2008qs,Kaloper:2008fb,Germani:2009iq,Kobayashi:2009hj, Koivisto:2009sd,Koivisto:2009ew,Koivisto:2009fb,Koivisto:2012xm,DeFelice:2012wy}. 
The coupling between the scalar $\phi$ and the 3-form field
generates the effective potential 
$V_{\rm eff} (\phi)$ for the scalar \cite{Dvali:2005an,Kaloper:2008qs}.   
This mechanism was used for the realization of chaotic inflation 
by the mixing between an axion and the 3-form \cite{Kaloper:2008fb}. 
In a more general setup with multiple scalar and axions,  
the interactions with 3-forms were studied at length in 
Refs.~\cite{Bielleman:2015ina, Ibanez:2015fcv, Valenzuela:2016yny} 
(see also Ref.~\cite{Farakos:2017jme}).
A crucial property exploited in those references rests on the fact that 
the 3-form field strength $F_{\mu \nu \lambda \rho}$ is proportional to the volume 
element in 4-dimensional space-time, so it can be regarded as an effective cosmological constant term. 
In this regard, such couplings are not only relevant for inflation but also for 
the dynamics of dark energy.

Recently, three of the present authors \cite{Almeida:2018fwe} performed a systematic construction of 
gauge-invariant theories with coupled $p$-forms on a $D$-dimensional background. 
The construction was restricted to up to first-order derivatives of fields 
in the action, so the theories do not contain, for instance, the 
second-order derivatives typical from the Galileon class of
interactions \cite{Fairlie1,Fairlie2,Nicolis:2008in,Deffayet:2009wt}. 
Indeed, even if we include the second-order derivatives of $p$-form fields 
in the action, there is a no-go theorem stating that Galilean interactions 
for odd $p$-forms preserving gauge invariance and Lorentz 
symmetry are forbidden\footnote{If we break the $U(1)$ gauge symmetry, it is possible to construct massive vector-tensor theories 
with Galileon-type derivative 
interactions \cite{Heisenberg,Tasinato1,Tasinato2,Allys,Jimenez2016,Heisenberg:2016eld,Allys:2016jaq}.} \cite{Deffayet:2010zh,Deffayet:2017eqq}. 
In Ref.~\cite{Almeida:2018fwe}, the authors derived the equations of motion 
on the homogenous and isotropic background for coupled $p$-form theories 
and applied them to the dynamics of dark energy in the presence of 
interactions between 0- and 3-forms.

If 1- and 2-forms are present besides 0- and 3-forms, we need to consider 
the Bianchi-type cosmological background to discuss the fate 
of anisotropies \cite{Normann:2017aav}. 
In this paper, we study the dynamics of anisotropic inflation for the most general 
parity-invariant coupled $p$-form theories with first-order derivatives of fields 
in the action. For the system in which the scalar field $\phi$ is independently 
coupled to 1- or 2-form fields, the anisotropic hair can survive during 
slow-roll inflation. If 1- and 2-form fields coexist, it was shown in 
Ref.~\cite{Ito} that anisotropic inflation supported by their couplings with 
the scalar can occur for the exponential potential of $\phi$. 
This case corresponds to anisotropic power-law inflation \cite{Kanno:2010nr,Ohashi:2013pca}, 
so there is no exit to the subsequent reheating stage. Instead, we will perform 
a more general analysis without specifying the scalar potential 
and derive conditions under which the anisotropic shear can be
supported by both 1- and 2-form fields.

Our study is general enough to cover the 
interaction between 1- and 2-form fields 
(as advocated in Refs.~\cite{Blau:1989bq,Blau:1989dh,Horowitz:1989ng,Allen:1990gb,Quevedo:1996uu,Dvali:2005ws,Birmingham:1991ty}).
Indeed, we show the existence of new anisotropic inflationary 
solutions along which the 2-form energy density is supported 
by the coupling with the 1-form.
We obtain analytic formulas of the anisotropic shear to 
the Hubble expansion rate for general inflaton potentials. 
The survival of anisotropic hair will be also numerically confirmed 
for a concrete inflaton potential.

This paper is organized as follows. In section \ref{sec:syspf} we briefly review 
the basics of $p$-forms by 
following the general results of Ref.~\cite{Almeida:2018fwe}. 
In section \ref{unsec} we study the uncoupled system between 1- and 2-forms and find general conditions 
under which the anisotropic shear survives during inflation. 
In section \ref{coupledbf} we take into account the interaction
between 1- and 2-forms and obtain a new class of anisotropic inflationary 
solutions. Section \ref{sec:Conclusions} is devoted to conclusions.

Throughout the paper, we use the Lorentzian metric $g_{\mu\nu}$ with the sign 
convention $(-,+,+,+)$. Greek indices $\alpha, \beta, \gamma\cdots $ 
denote space-time coordinates, while latin indices $ i, j, k,\cdots$ 
represent spatial coordinates.

\section{Coupled \texorpdfstring{$\boldsymbol{p}$}{p}-forms and anisotropic cosmological background}
\label{sec:syspf}

The general gauge-invariant action of coupled $p$-forms was derived in 
Ref.~\cite{Almeida:2018fwe} by restricting the derivatives of fields 
up to first order. In Sec.~\ref{actionsec}, we briefly review such a system 
and show how the 3-form coupled to the scalar field $\phi$ generates 
the effective potential for $\phi$. 
In Sec.~\ref{sec:background}, we derive the field equations of motion 
on the anisotropic cosmological background for parity-invariant coupled 
$p$-form theories.

\subsection[Action of coupled $p$-forms]{\boldmath Action of coupled $p$-forms}
\label{actionsec}

In the 4-dimensional space-time, the $p$-forms $A_{(p)}$, 
where $p=1,2,3$, are defined, respectively, by 
\ba\label{pnot} 
& &
A_{(1)} = A_{(1) \, \mu_{1}   } \mr{d}x^{\mu_{1}},  \qquad
A_{(2)} = \frac{1}{2}A_{(2) \, \mu_{1} \mu_{2}  } \mr{d}x^{\mu_{1}} \wedge \mr{d}x^{\mu_{2}},  \nonumber \\
& &
A_{(3)} = \frac{1}{6}A_{(3) \, \mu_{1} \mu_{2} \mu_{3}  } \mr{d}x^{\mu_{1}} \wedge \mr{d}x^{\mu_{2}} \wedge \mr{d}x^{\mu_{3}}\,,
\ea
where $A_{(2)\mu_1 \mu_2}$ and $A_{(3)\mu_1 \mu_2 \mu_3}$ are totally anti-symmetric tensors, 
and $\wedge$ represents the wedge product.
Associated with them, we define the  field strengths,
\be
F_{(1)  \mu_{1}  \mu_{2}  } = 2 \partial_{[\mu_1}A_{(1)\mu_2]},\qquad
F_{(2)  \mu_{1} \mu_{2}  \mu_{3}} = 3 \partial_{[\mu_1}A_{(2) \mu_2 \mu_3]},\qquad 
F_{(3) \, \mu_{1} \mu_{2} \mu_{3}  \mu_{4}  } = 
4 \partial_{[\mu_1}A_{(3) \mu_2 \mu_3 \mu_4]}, \quad
\ee
and the Hodge duals,
\ba
& &
\tilde{F}_{ (1)  \mu_{1}  \mu_{2} } = \frac{1}{2} {\cal E}_{ \mu_{1} \mu_{2} \mu_{3} \mu_{4} }F_{(1)}{}^{\mu_{3} \mu_{4}}, \qquad 
\tilde{ F}_{ (2) \mu_{1}  } = \frac{1}{6} {\cal E}_{ \mu_{1} \mu_{2} \mu_{3} \mu_{4} } F_{(2)}{}^{ \mu_{2} \mu_{3} \mu_{4} },\nonumber \\
& &
\tilde{ F}_{ (3) } = \frac{1}{24} {\cal E}_{ \mu_{1} \mu_{2} \mu_{3} \mu_{4} } F_{(3)}{}^{ \mu_{1} \mu_{2} \mu_{3} \mu_{4} }\,,
\ea
where  ${\cal E}_{\mu_1 \mu_2 \mu_3 \mu_4}
=\sqrt{-g} \,\epsilon_{\mu_1 \mu_2 \mu_3 \mu_4}$, 
$g$ is the determinant of metric tensor $g_{\mu \nu}$,  
and $\epsilon_{\mu_1 \mu_2 \mu_3 \mu_4}$ is the 
Levi-Civita symbol with the convention
$\epsilon_{0123}=1$.

The Lagrangian of interacting $p$-forms consistent with Abelian gauge 
invariance was derived in Ref.~\cite{Almeida:2018fwe} by restricting the 
field derivatives up to first order.
Besides $p$-forms, we take into account a canonical scalar field $\phi$ 
with the potential $V(\phi)$. 
For the gravity sector we consider the Einstein-Hilbert Lagrangian $M_{\rm pl}^2 R/2$, 
where $M_{\rm pl}$ is the reduced Planck mass and $R$ is the Ricci scalar. 
We assume that the scalar and form fields are minimally coupled to gravity.
Then, the total action of such a system is given by  
\be
{\cal S}=\int {\rm d}^4 x \sqrt{-g}\,
\left( \frac{M_{\rm pl}^2}{2}R
+{\cal L}_{\phi}+
{\cal L}_{p} \right)\,,
\label{MandT}
\ee
where
\ba
{\cal L}_{\phi}
&=& -\frac{1}{2}\partial_{\mu}\phi \partial^{\mu}\phi 
-V(\phi)\,,\\
{\cal L}_{p} 
&=& {\cal L}_{M} + {\cal L}_{T}\,.
\ea
The Lagrangian ${\cal L}_{p}$, which arises from $p$-forms, 
consists of two contributions:
\be
\label{FstarF}
{\cal L}_{M}
= -\frac{1}{2} \sum_{p=1}^{3} \frac{f_{p}(\phi)}{(p+1)!} 
F_{(p)}^2\,, \qquad 
{\cal L}_{T} = - g(\phi) F_{(1)}\wedge  F_{(1)} 
+ \sum_{p=0}^{3} h_{p}(\phi) A_{(p)} \wedge  F_{(3-p)}\,,
\ee
where $f_{p}(\phi)$, $g(\phi)$, and $h_{p}(\phi)$ are 
$\phi$-dependent functions, and 
\be
\label{F2}
F_{(p)}^2\equiv F_{(p)  \mu_1\mu_2\ldots \mu_{p+1}}F_{(p)}{}^{\mu_1\mu_2\ldots \mu_{p+1}}\,.
\ee
In the sum in ${\cal L}_{T}$ we extended the notation \eqref{pnot} to include the scalar field as a $0$-form, 
that is,  $A_{(0)} = \phi$ and 
$F_{(0)\mu}=\partial_{\mu} \phi$. 
The Lagrangian ${\cal L}_{M}$ corresponds to dynamical, 
quadratic, Maxwell-like terms, whereas ${\cal L}_{T}$ 
represents topological terms. 
The contribution $g(\phi) F_{(1)}\wedge  F_{(1)}$ to 
${\cal L}_{T}$ is the Chern-Pontyiagin 
term (or $\theta$-term), which affects the background dynamics 
when it is coupled to the scalar field through the function $g(\phi)$. 
The other contributions to ${\cal L}_{T}$ are the so called $BF$-terms \cite{Blau:1989dh,Horowitz:1989ng,Blau:1989bq,Allen:1990gb,Quevedo:1996uu,Dvali:2005ws,Birmingham:1991ty}. 
In total, we have four terms of this type: 
\ba
& A_{(0)} \wedge  F_{(3)}  \propto  \phi \tilde{F}_{(3)}, \qquad  & A_{(1)} \wedge  F_{(2)}  \propto  A_{(1)\mu}  \tilde{F}_{(2)}{}^{ \mu}, \nonumber \\
& A_{(2)} \wedge  F_{(1)}  \propto  A_{(2)\mu_1 \mu_2}  
\tilde{F}_{(1)}{}^{ \mu_1 \mu_2}, \qquad 
& A_{(3)} \wedge  F_{(0)}  \propto  \tilde{A}_{(3)\mu} \partial^{\mu}\phi\,.
\ea
After integration by parts, however, it can be seen that $A_{(0)} \wedge  F_{(3)}$ and 
$A_{(1)} \wedge  F_{(2)}$ are equivalent to $A_{(3)} \wedge  F_{(0)}$ and 
$A_{(2)} \wedge  F_{(1)}$, respectively.
In the following, we will keep the terms 
$A_{(0)} \wedge  F_{(3)}$ and  $A_{(2)} \wedge  F_{(1)}$.
After redefining the coupling functions, 
the complete $p$-form Lagrangian is expressed as
\be
{\cal L}_{p}= -\frac{1}{2} \sum_{p=1}^{3} \frac{f_{p}(\phi)}{(p+1)!}F_{(p)}^2
- \frac{g_1(\phi)}{4}  F_{(1) \mu_{1} \mu_{2} }  \tilde{F}_{(1)}{}^{ \mu_{1} \mu_{2} } 
- \frac{g_2(\phi) }{2}A_{(2)}{}_{\mu_1 \mu_2}  \tilde{F}_{(1)}{}^{ \mu_{1} \mu_{2}  }  - {g_3(\phi)}  \tilde{F}_{(3)}\,.
\label{eq:lphiAp4}
\ee
We note that further couplings between $1$-, $2$-, and $3$- forms are already included 
in the previous Lagrangian. For instance, we have
\ba
{F}_{(3)}{}^{ \mu_1 \mu_2 \mu_3 \mu_4} F_{(1)\mu_1 \mu_2}  F_{(1)\mu_3 \mu_4} 
&\propto & 
F_{(1)\mu_1 \mu_2}  \tilde{F}^{(1)\mu_1 \mu_2}\,,\notag\\ 
{F}_{(3)}{}^{ \mu_1 \mu_2 \mu_3 \mu_4} \tilde{F}_{(1)\mu_1 \mu_2}  F_{(1)\mu_3 \mu_4} 
&\propto& 
F_{(1)\mu_1 \mu_2}  {F}^{(1)\mu_1 \mu_2}, \nonumber \\
{F}_{(3)}{}^{ \mu_1 \mu_2 \mu_3 \mu_4} F_{(2)\mu_1 \mu_2 \mu_3}  \tilde{F}_{(2) \mu_4} 
& \propto &  
\tilde{F}^{(2) \mu_4}  \tilde{F}_{(2) \mu_4} \propto   {F}_{(2) \mu_1 \mu_2 \mu_3}  {F}^{(2) \mu_1 \mu_2 \mu_3}\,.
\ea
Other possible combinations like
\ba
{F}_{(2)}{}^{ \mu_1 \mu_2 \mu_3 } \tilde{F}_{(2)\mu_1 }  F_{(1)\mu_2 \mu_3} \propto {\cal E} ^{\sigma \mu_1 \mu_2 \mu_3  } 
\tilde{F}_{(2)\sigma }  \tilde{F}_{(2)\mu_1 }  F_{(1)\mu_2 \mu_3} \propto F_{(3)}{}^{\sigma \mu_1 \mu_2 \mu_3  } \tilde{F}_{(2)\sigma } 
 \tilde{F}_{(2)\mu_1 }  F_{(1)\mu_2 \mu_3} =0\,, \nonumber \\ 
{F}_{(2)}{}^{ \mu_1 \mu_2 \mu_3 } \tilde{F}_{(2)\mu_1 }  \tilde{F}_{(1)\mu_2 \mu_3} \propto 
{\cal E} ^{\sigma \mu_1 \mu_2 \mu_3  } \tilde{F}_{(2)\sigma }  \tilde{F}_{(2)\mu_1 }  
\tilde{F}_{(1)\mu_2 \mu_3} \propto F_{(3)}{}^{\sigma \mu_1 \mu_2 \mu_3  } \tilde{F}_{(2)\sigma } 
 \tilde{F}_{(2)\mu_1 }  \tilde{F}_{(1)\mu_2 \mu_3} =0\, , \nonumber \\
\ea
are null due to contractions between antisymmetric and symmetric indices. 
Moreover, couplings involving the kinetic term $X =-\partial_{\mu}\phi \partial^{\mu}\phi/2$ of the scalar, 
such as $f(\phi, X)F_{(p)}^2$ or 
$\partial_{\mu} \phi F^{\mu \alpha} \partial_{\nu}\phi F^{\nu}{}_{\alpha}$, 
suffer from Hamiltonian instability and non-hyperbolicity of the 
equations of motion \cite{Fleury2014qfa}, so, we will not consider 
those terms either. 

Now, we will explicitly show that, after integrating out the $3$-form field from 
the action, it can be absorbed into the scalar potential. 
To see this, we isolate the terms in ${\cal L}_{\phi}+{\cal L}_{p}$ 
corresponding to the scalar field and the $3$-form, as
\be
{\cal L}_{\phi A_{(3)}} = 
-\frac{1}{2}\partial_{\mu}\phi \partial^{\mu}\phi  - V(\phi) -  \frac{f_{3}(\phi)}{48}F_{(3)}^2 -  \frac{{g_3(\phi)}}{24} \frac{\epsilon^{\mu\nu\sigma\rho}}{\sqrt{-g}} {F}_{(3) \mu\nu\sigma\rho }\,.
\label{phi3f}
\ee
Varying the corresponding action with respect to $\phi$ 
and $A_{(3)}$ respectively, it follows that 
\ba
\Box \phi -V_{,\phi}-\frac{f_{3,\phi}}{48} F_{(3)}^2   -  \frac{g_{3,\phi}}{24}   \frac{\epsilon^{\mu\nu\sigma\rho}}{\sqrt{-g}} {F}_{(3) \mu\nu\sigma\rho } 
&=& 0\,,
\label{scalareq}\\
\nabla^{\mu}\left[f_3 F_{(3) \mu \nu \sigma \rho} 
+ g_{3} \sqrt{-g} \epsilon_{\mu \nu \sigma \rho} \right] 
&=& 0\,,
\label{3feom}
\ea
where the notation $V_{,\phi}={\rm d}V/{\rm d}\phi$ is used.
We exploit the fact that $F_{(3) \mu \nu \sigma \rho}$ 
is proportional to the volume element $\sqrt{-g}$, i.e., 
\be
F_{(3) \mu \nu \sigma \rho}=
X(x^{\mu}) \sqrt{-g} \epsilon_{\mu \nu \sigma \rho}\,,
\label{F3so}
\ee
where $X(x^{\mu})$ is a scalar function.
Substituting Eq.~(\ref{F3so}) into \eqref{3feom}, we obtain
\be
X(x^{\mu}) = \frac{c-g_{3}(\phi) }{f_{3}(\phi)}\,,
\label{X4}
\ee
where $c$ is an integration constant. 
Plugging this solution into Eq.~\eqref{phi3f} and using the properties 
$F_{(3)}^2 = -4! X^2$ and $\epsilon^{\mu\nu\sigma\rho}
{F}_{(3) \mu\nu\sigma\rho}/\sqrt{-g} = -4! X$, 
the Lagrangian ${\cal L}_{\phi A_{(3)}}$ reduces to 
\be
{\cal L}_{\phi A_{(3)}} = 
-\frac{1}{2}\partial_{\mu}\phi \partial^{\mu}\phi 
-V (\phi) +\frac{c^2 - g_{3}(\phi)^2}{2f_{3}(\phi)}\,.
\ee
Varying the corresponding action with respect to $\phi$ and comparing the resulting 
equation of motion with Eq.~(\ref{scalareq}), we find that the 
constant $c$ is fixed to zero. 
Then, the coupled system of scalar and $3$-forms is equivalent to 
the Lagrangian, 
\be
{\cal L}_{\phi A_{(3)}} = 
-\frac{1}{2}\partial_{\mu}\phi \partial^{\mu}\phi 
-\Ve(\phi)\,, 
\ee
where $\Ve(\phi)$ is the effective potential given by 
\be
\Ve(\phi)  \equiv V (\phi) 
+\frac{g_{3}(\phi)^2}{2f_{3}(\phi)}\,.
\ee
Thus, in the presence of coupling functions $f_3(\phi)$ and $g_3(\phi)$, 
the $3$-form induces the potential $g_3(\phi)^2/[2f_3(\phi)]$ 
for the scalar field.

Keeping in mind that we are considering a gauge-invariant theory, 
the coupling $g_2(\phi)$ in Eq.~(\ref{eq:lphiAp4}) is constrained to be constant. 
We will use the notation: 
\be
g_2(\phi)=m_{v}={\rm constant}\,.
\ee
Additionally, in this work, we focus on parity-conserving 
theories invariant under the transformation 
${\cal P}$: $\vec{x} \to -\vec{x}$.
Hence we set the coupling $g_1(\phi)$ to 0 
in the following.
Despite the presence of the Levi-Civita symbol 
in $A_{(2)}{}_{\mu_1 \mu_2}  
\tilde{F}_{(1)}{}^{ \mu_{1} \mu_{2}}$, this can be absorbed as a parity conserving longitudinal mass term for the 1-form $A_{(1) \mu}$ \cite{Allen:1990gb,Quevedo:1996uu,Dvali:2005ws, Almeida:2018fwe}\footnote{The other way to see this is to use the equation of motion for the 2-form field, which implies $F_{(1)}{}^{\mu\nu} = -1/(3g_2)\nabla^{[\mu}f_2 \tilde{F}_{(2)}{}^{\nu]}$. Replacing $F_{(1)}{}^{\mu\nu}$ back into the Lagrangian, it can be seen that there are two 
Levi-Civita symbols in the  $A_{(1)} \wedge  F_{(2)}$ 
term, and hence the Lagrangian is parity-invariant on shell.}. 
Then, the Lagrangian (\ref{eq:lphiAp4}) preserving gauge symmetry 
and parity invariance yields
\be
\mL{L}_{p}= -\frac{1}{2}\partial_{\mu}\phi \partial^{\mu}\phi  - \Ve(\phi)   - \frac{1}{2} \sum_{p=1}^{2} \frac{f_{p}(\phi)}{(p+1)!}F_{(p)}^2 
- \frac{1}{2}m_v\,A_{(2)}{}_{\mu_1 \mu_2}  
\tilde{F}_{(1)}{}^{ \mu_{1} \mu_{2}}\,.
\label{eq:LTp}
\ee
For the later convenience, we will use the following notations 
for the $p$-forms and their respective field strengths:
\be
A_{(1)\mu}  \equiv  A_{\mu}\,, \qquad 
A_{(2)\mu_1 \mu_2} \equiv B_{\mu_1 \mu_2}\,, \qquad
F_{(1)\mu_1 \mu_2}  \equiv  F_{\mu_1 \mu_2}\,, \qquad 
F_{(2)\mu_1 \mu_2 \mu_3} \equiv H_{\mu_1 \mu_2 \mu_3}\,, 
\ee
and 
\be
F^2 \equiv F_{ \mu_1 \mu_2}F^{ \mu_1 \mu_2}\,,\qquad
H^2  \equiv H_{ \mu_1 \mu_2 \mu_3}H^{ \mu_1 \mu_2 \mu_3}\,,\qquad
B \tilde{F} \equiv B_{\, \mu_1 \mu_2}\tilde{F}^{\mu_1\mu_2}\,.
\ee
Then, the total action (\ref{MandT}) reduces to 
\be 
\mL{S} = \int \mr{d}^4 x \sqrt{-g}  \left[  \frac{M_{\rm{pl}}^2}{2}R   -\frac{1}{2}\partial_{\mu}\phi \partial^{\mu}\phi  - \Ve(\phi)   - \frac{1}{4} f_{1}(\phi) F^2 -  \frac{1}{12} f_{2}(\phi) H^2 - \frac{1}{2} m_v B\tilde{F} \right]\,.
\label{eq:LT}
\ee

Varying the action (\ref{eq:LT}) with respect to $g_{\alpha \beta}$
on general curved backgrounds, it follows that 
\be
M_{\rm pl}^2 \left(
R_{\alpha\beta}-\frac{1}{2}R g_{\alpha\beta} \right)
=T_{\alpha\beta}\,,
\ee
where $R_{\alpha\beta}$ is the Ricci tensor, and 
$T_{\alpha\beta}$ is the energy-momentum tensor defined by 
\ba
T_{\alpha\beta} &=&
\partial_{\alpha}\phi \partial_{\beta}\phi 
- \frac{1}{2} g_{\alpha\beta} \partial_{\sigma}\phi \partial^{\sigma}\phi 
- g_{\alpha\beta}\Ve(\phi)  
+f_1(\phi)\left( F_{ \beta\gamma}F_{ \alpha}{}^{\gamma} - \frac{1}{4}  g_{\alpha\beta} F^2 \right) \nonumber \\
& &
+ f_2 (\phi) \left(\frac{1}{2} H_{\beta}{}^{\gamma\delta}H{}_{ \alpha\gamma\delta} - \frac{1}{12}g_{\alpha\beta}H^2 \right)\,.
\ea
The topological $B \tilde{F}$ term does not contribute to the energy-momentum 
tensor as they are independent of the metric.
The equations of motion for $\phi$, 1-form, and $2$-form 
following from the action (\ref{eq:LT}) are given, 
respectively, by  
\ba
\label{eq:EPhi}
\square \phi - \Vep -\frac{f_{1,\phi}}{4}F^2 - \frac{f_{2,\phi}}{12}  H^2 &=& 0\,, \\
\nabla^{\mu} \left[ f_1(\phi) F_{\mu\nu} +m_v  \tilde{B}_{\mu\nu}  \right]  &=& 0\,, \label{eq:EA}\\[1mm] 
 \nabla^{\mu}\left[ f_2(\phi) H_{\mu\nu \alpha}  \right]
 -m_v  \tilde{F}_{\nu\alpha}  &=& 0\,, \label{eq:EB}
\ea
where $\tilde{B}_{\mu\nu}={\cal E}_{\mu\nu\alpha\beta}B^{\alpha\beta}/2$.
These equations are complemented with the Bianchi identities,
\be
\nabla_{\mu}\tilde{F}^{\mu\nu} = 0, \qquad 
\nabla_{\mu}\tilde{H}^{\mu}=0\,,
\ee
together with the gauge choice. 

\subsection{Background equations on the anisotropic 
cosmological background}
\label{sec:background}

We derive the dynamical equations of motion for 
the reduced action \eqref{eq:LT} on the anisotropic cosmological background. 
We consider the 1-form field $A_{\mu}$ alined in the $x$ direction, 
such that $A_{\mu}=(0, v_A(t), 0, 0)$, where 
$v_A$ depends on the cosmic time $t$. 
For the 2-form field $B_{\mu \nu}$, we choose the configuration 
orthogonal to the 1-form field, such that 
$B_{\mu \nu} \mr{d}x^{\mu} \wedge \mr{d}x^{\nu}
=2v_B(t) \mr{d}y \wedge \mr{d}z$, where $v_B$ 
depends on $t$.
Since there is a rotational symmetry in the $(y,z)$ plane, 
we can take the line element in the following form, 
\be
\mr{d}s^2 = -N(t)^2 \mr{d}t^2 + e^{2\alpha(t)} \left[ e^{-4\sigma (t)}\mr{d}x^2
+e^{2\sigma (t)}(\mr{d}y^2+\mr{d}z^2) \right] \ ,
\label{anisotropic-metric}
\ee
where $N(t)$ is a lapse function, $a \equiv e^{\alpha(t)}$ is an isotropic 
scale factor, and $\sigma(t)$ is a spatial shear.
The non-vanishing components $F_{\mu \nu}$ and $B_{\mu \nu}$ 
are given by $F_{01}=-F_{10}=\dot{v}_A$ and 
$B_{23}=-B_{32}=v_B$, where a dot represents 
a derivative with respect to $t$. 
Then, the action \eqref{eq:LT} yields
\ba
{\cal S}&=& \int {\rm d}^4 x \left[ \frac{3M_{\rm pl}^2 
e^{3\alpha}}{N} \left(\dot{\sigma}^2 -\dot{\alpha}^2 
\right)+e^{3\alpha} \frac{\dot{\phi}^2}{2N} 
-Ne^{3\alpha} V_{\rm eff}(\phi)
+\frac{f_1(\phi)}{2N}e^{\alpha+4\sigma} \dot{v}_A^2 \right.
\nonumber \\
& &\qquad \quad
\left. +\frac{f_2(\phi)}{2N}e^{-\alpha-4\sigma} \dot{v}_B^2
+m_v \dot{v}_A v_B \right]\,.
\label{ac2}
\ea

Varying the action (\ref{ac2}) with respect to $v_A$ and $v_B$, 
respectively, and setting $N=1$ at the end, it follows that 
\ba
\frac{\mr{d}}{\mr{d} t} \left[ f_1(\phi) e^{\alpha + 4 \sigma}\dot{v}_A + m_v v_B \right] &= 0\,, 
\quad \Rightarrow \quad  f_1(\phi) e^{\alpha + 4 \sigma}\dot{v}_A + m_v v_B  = p_A\,, \label{be4} \\[2mm]
\frac{\mr{d}}{\mr{d} t} \left[ f_2(\phi) e^{-\alpha - 4 \sigma}\dot{v}_B - m_v v_A \right] &= 0\,, 
\quad \Rightarrow \quad f_2(\phi) e^{-\alpha - 4 \sigma}\dot{v}_B - m_v v_A = p_B\,, \label{be5} 
\ea
where $p_A$ and $p_B$ are constants. 
Expanding the time derivatives of 
Eqs.~(\ref{be4}) and (\ref{be5}), we find
\ba
\ddot{v}_A+\left( \frac{f_{1,\phi}}{f_1} 
\dot{\phi}+ \dot{\alpha}+4\dot{\sigma} 
\right)\dot{v}_A+\frac{m_v^2}{f_1 f_2} \left( v_A
+\frac{p_B}{m_v} \right)=0\,,\label{vAeq}\\
\ddot{v}_B+\left( \frac{f_{2,\phi}}{f_2} 
\dot{\phi} -\dot{\alpha}-4\dot{\sigma} 
\right)\dot{v}_B+\frac{m_v^2}{f_1 f_2} \left( v_B
-\frac{p_A}{m_v} \right)=0\,.\label{vBeq}
\ea
This means that both 1- and 2-form fields acquire 
the effective mass term $m_v/\sqrt{f_1 f_2}$ 
through their interactions. 

Varying the action (\ref{ac2}) with respect to 
$N, \alpha, \sigma$, and $\phi$, we obtain
\begin{align}
3M_{\rm pl}^2 \left( \dot{\alpha}^2-\dot{\sigma}^2 \right)
&=\frac{1}{2} \dot{\phi}^2+V_{\rm eff}(\phi)+\rho_A+\rho_B\,,
\label{eq:be1} \\
M_{\rm pl}^2 \left( \ddot{\alpha}+3\dot{\alpha}^2 \right)
&=V_{\rm eff}(\phi)+\frac{1}{3} \rho_A+\frac{2}{3} \rho_B\,,
\label{eq:be2} \\
M_{\rm pl}^2 \left( \ddot{\sigma}+3 \dot{\alpha}\dot{\sigma} \right)
&=\frac{2}{3} \rho_A-\frac{2}{3} \rho_B\,,
\label{eq:be3} \\
\ddot{\phi}+3 \dot{\alpha} \dot{\phi}+V_{{\rm eff},\phi}&
-\frac{f_{1,\phi}}{f_1} \rho_A-\frac{f_{2,\phi}}{f_2} \rho_B=0\,,
\label{eq:be6}
\end{align}
where $\rho_A$ and $\rho_B$ are the energy densities 
of 1- and 2-forms defined, respectively, by 
\be
\rho_A=\frac{f_1(\phi)}{2} e^{-2\alpha+4\sigma} \dot{v}_A^2\,,\qquad 
\rho_B=\frac{f_2(\phi)}{2} e^{-4\alpha-4\sigma} \dot{v}_B^2\,.
\label{rhoAB}
\ee
On using Eqs.~(\ref{be4})-(\ref{be5}) and 
(\ref{vAeq})-(\ref{vBeq}), (\ref{rhoAB}), the energy 
densities $\rho_A$ and $\rho_B$ obey  
\ba
& &
\dot{\rho}_A=-4\rho_A \left( \dot{\alpha}+\dot{\sigma}
+ \frac{\dot{f}_{1}}{4f_1} 
\right)
-2m_v \sqrt{\frac{\rho_A \rho_B}{f_1 f_2}}\,,
\label{drhoA}\\
& &
\dot{\rho}_B=-2\rho_B \left( \dot{\alpha}-2\dot{\sigma}
+ \frac{\dot{f}_{2}}{2f_2} 
 \right)
+2m_v \sqrt{\frac{\rho_A \rho_B}{f_1 f_2}}\,, 
\label{drhoB}
\ea
where we assumed that $\dot{v}_A>0$ and $\dot{v}_B>0$ without loss of generality. 
In subsequent sections, we exploit the above background 
equations of motion to study the dynamics of anisotropic inflation.

\section{Anisotropic inflation for uncoupled 1-form and 2-form fields}
\label{unsec}

It is known that anisotropic inflation can be realized by 
the couplings of the scalar field $\phi$ with 
1-form \cite{Watanabe} or 2-form \cite{Ohashi1} fields. 
We study the dynamics of anisotropic inflation driven by 
both 1- and 2-form fields without specifying 
the form of the effective scalar potential $V_{\rm eff}(\phi)$.
In this section, we consider the system in which 
the 1-and 2-form fields are uncoupled to each other, i.e., 
\be
m_v=0\,.
\ee
In this case, we obtain 
$\dot{v}_A=p_A f_1^{-1}(\phi) e^{-\alpha-4\sigma}$ and 
$\dot{v}_B=p_B f_2^{-1}(\phi) e^{\alpha+4\sigma}$ 
from Eqs.~(\ref{be4}) and (\ref{be5}).
Then, the 1- and 2-form energy densities 
in Eq.~(\ref{rhoAB}) reduce to 
\be
\rho_A=\frac{p_A^2}{2f_1(\phi)} e^{-4\alpha-4\sigma}\,,\qquad 
\rho_B=\frac{p_B^2}{2f_2(\phi)} e^{-2\alpha+4\sigma}\,.
\label{rhoAB2}
\ee
Provided that  $|\sigma| \ll \alpha$, the couplings 
$f_1(\phi) \propto e^{-4\alpha}$ and 
$f_2(\phi) \propto e^{-2\alpha}$ lead to 
nearly constant values of $\rho_A$ and $\rho_B$, respectively. 
During inflation in which the scalar field $\phi$ evolves slowly 
along the nearly flat potential $V_{\rm eff}(\phi)$, 
we can resort to the slow-roll 
approximations $\dot{\phi}^2/2 \ll V_{\rm eff}(\phi)$ 
and $|\ddot{\phi}| \ll |3 \dot{\alpha} \dot{\phi}|$.
Ignoring also the contributions of 1- and 2-form fields to 
Eqs.~(\ref{eq:be1}) and (\ref{eq:be6}), we obtain
$3M_{\rm pl}^2 \dot{\alpha}^2 \simeq V_{\rm eff}(\phi)$ and 
$3 \dot{\alpha} \dot{\phi} \simeq -V_{{\rm eff}, \phi}$, 
so that $\mr{d}\alpha/\mr{d}\phi \simeq 
-\Ve/(M_{\rm pl}^2 \Vep)$.
Then, the critical couplings 
$f_1(\phi) \propto e^{-4\alpha}$ and $f_2(\phi) \propto e^{-2\alpha}$ 
correspond, respectively, to  
\be
f_1(\phi)=e^{ 4 \int\frac{\Ve}{M_{\rm pl}^2 \Vep}
\mr{d}\phi}\,,\qquad 
f_2 (\phi)=e^{ 2 \int\frac{\Ve}{M_{\rm pl}^2 \Vep}
\mr{d}\phi}\,.
\ee
In Refs.~\cite{Watanabe,Ohashi1,Ohashi2}, it was shown that 
these couplings separately give rise to anisotropic inflation 
with the non-vanishing shear 
$\Sigma \equiv \dot{\sigma}$ satisfying
\ba
\frac{\Sigma}{H} \simeq \frac{\epsilon}{12 (\alpha+\alpha_0)}\,,\qquad 
{\rm for} \quad p_A \neq 0\,,\quad p_B=0\,,\\
\frac{\Sigma}{H} \simeq -\frac{\epsilon}{3 (\alpha+\alpha_0)}\,,\qquad 
{\rm for} \quad p_A=0\,,\quad p_B \neq 0\,,
\ea
where $\alpha_0$ is a constant, and 
\be
H \equiv \dot{\alpha}\,,\qquad 
\epsilon \equiv -\frac{\dot{H}}{H^2}\,.
\ee

It is also possible to sustain the anisotropic hair 
for the functions given by 
\be
f_1(\phi)=e^{ 4c_1 \int\frac{\Ve}{M_{\rm pl}^2 \Vep}
\mr{d}\phi}\,,\qquad 
f_2 (\phi)=e^{ 2c_2 \int\frac{\Ve}{M_{\rm pl}^2 \Vep}
\mr{d}\phi}\,,
\label{f12}
\ee
where $c_1$ and $c_2$ are constants.
In the following, we study the dynamics of anisotropic inflation 
for the system in which the couplings 
of the form (\ref{f12}) coexist. 
The similar analysis was carried out in Ref.~\cite{Ito} for the 
exponential potential $\Ve(\phi)=V_0e^{\lambda \phi/M_{\rm pl}}$, 
in which case the couplings (\ref{f12}) correspond to 
$f_1(\phi)=e^{4c_1 \phi/(\lambda M_{\rm pl})}$ and 
$f_2(\phi)=e^{2c_2 \phi/(\lambda M_{\rm pl})}$. 
In this case there is no exit from the inflationary stage,  
so we will perform a more general treatment without 
assuming the form of $\Ve(\phi)$.

If the contributions of 1- and 2-form fields to Eq.~(\ref{eq:be6}) 
are negligibly small, there is the approximation relation 
$\mr{d}\alpha/\mr{d}\phi \simeq -\Ve/(M_{\rm pl}^2 \Vep)$. 
In this regime, the couplings evolve as 
$f_1(\phi) \propto e^{-4c_1 \alpha}$ and 
$f_2(\phi) \propto e^{-2c_2 \alpha}$, so that 
$\rho_A \propto e^{4(c_1-1)\alpha-4\sigma}$ and 
$\rho_B \propto e^{2(c_2-1)\alpha+4\sigma}$. 
Provided that $|\sigma| \ll \alpha$, $\rho_A$ and 
$\rho_B$ increase for $c_1>1$ and $c_2>1$, respectively.
Then, the energy densities of 1- and 2-form fields should
start to contribute to the background cosmological dynamics. 

In the following, we perform a more refined treatment for the 
dynamics of anisotropic inflation without 
ignoring the couplings of $\phi$ with 1- and 2-form fields 
in Eq.~(\ref{eq:be6}).
Applying the slow-roll approximation 
$|\ddot{\phi}| \ll |3 \dot{\alpha} \dot{\phi}|$ to Eq.~(\ref{eq:be6}), 
it follows that 
\be
\frac{\mr{d}\phi}{\mr{d} \alpha} \simeq -\frac{M_{\rm pl}^2 \Vep}{\Ve} 
\left( 1-\frac{c_1 p_A^2}{\epsilon_V \Ve} 
e^{-4\alpha-4\sigma-4c_1 \int \frac{\Ve}
{M_{\rm pl}^2 \Vep}\mr{d}\phi} 
-\frac{c_2 p_B^2}{2\epsilon_V \Ve} 
e^{-2\alpha+4\sigma-2c_2 \int \frac{\Ve}
{M_{\rm pl}^2 \Vep}\mr{d}\phi} 
\right)\,,
\label{dal1}
\ee
where 
\be
\epsilon_V \equiv \frac{M_{\rm pl}^2}{2}
\left( \frac{\Vep}{\Ve} \right)^2\,.
\ee
The evolution of inflaton $\phi$ slows down by the existence of 
couplings $f_1(\phi)$ and $f_2(\phi)$. 
Then, the solutions eventually enter the stage in which 
either $\rho_A$ or $\rho_B$ approaches a constant value 
smaller than $\Ve$. 
To understand this behavior, we first write 
Eq.~(\ref{dal1}) in the form 
\be
\frac{\mr{d}\alpha}{\mr{d}\phi}+\frac{\Ve}{M_{\rm pl}^2\Vep}
=\left( \frac{c_1 p_A^2}{\epsilon_V \Ve} 
e^{-4\alpha-4\sigma-4c_1 \int \frac{\Ve}
{M_{\rm pl}^2 \Vep}\mr{d}\phi} 
+\frac{c_2 p_B^2}{2\epsilon_V \Ve} 
e^{-2\alpha+4\sigma-2c_2 \int \frac{\Ve}
{M_{\rm pl}^2 \Vep}\mr{d}\phi} 
\right) \frac{\mr{d}\alpha}{\mr{d}\phi}\,.
\label{dal2}
\ee
Introducing the quantity 
\be
x \equiv e^{2c_1 \alpha+2c_1 \int \frac{\Ve}
{M_{\rm pl}^2 \Vep}\mr{d}\phi}\,,
\ee
Eq.~(\ref{dal2}) can be further expressed as
\be
\frac{\mr{d}x}{\mr{d}\phi}=\frac{c_1}{\epsilon_V \Ve} 
\left[ 2c_1p_A^2 e^{4(c_1-1)\alpha-4\sigma}x^{-1}
+c_2 p_B^2 e^{2(c_2-1) \alpha+4\sigma}
x^{1-c_2/c_1} \right] \frac{\mr{d}\alpha}{\mr{d}\phi}\,.
\label{dal3}
\ee
In what follows, we integrate Eq.~(\ref{dal3}) under the 
approximation that the quantity $\epsilon_V \Ve$ is constant.
We also neglect the time dependence of 
$\sigma$ under the condition that $|\sigma| \ll \alpha$.

When $p_B=0$, the integrated solution to Eq.~(\ref{dal3}) 
is given by 
\be
x^2=\frac{c_1^2p_A^2}{(c_1-1)\epsilon_V \Ve}
e^{4(c_1-1)\alpha-4\sigma}+\Sigma_A\,,
\label{x2}
\ee
where $\Sigma_A$ is a constant. 
For $c_1>1$, the first term on the right hand side of 
Eq.~(\ref{x2}) exponentially increases during inflation. 
Hence the quantity $x$ approaches the value
\be
x_A=\sqrt{\frac{c_1^2p_A^2}{(c_1-1)\epsilon_V \Ve}} 
e^{2(c_1-1)\alpha-2\sigma}\,.
\label{xA}
\ee
If $p_A=0$, then the solution to Eq.~(\ref{dal3}) reads
\be
x^{c_2/c_1}=\frac{c_2^2 p_B^2}{2(c_2-1) \epsilon_V \Ve} 
e^{2(c_2-1)\alpha+4\sigma}+\Sigma_B\,,
\ee
where $\Sigma_B$ is a constant. 
For $c_2>1$, the variable $x$ eventually approaches the value 
\be
x_B=\left[ \frac{c_2^2 p_B^2}{2(c_2-1) \epsilon_V \Ve} 
\right]^{c_1/c_2} e^{[2(c_2-1)\alpha+4\sigma]c_1/c_2}\,.
\label{xB}
\ee
Ignoring the $\sigma$-dependent terms in Eqs.~(\ref{xA}) and 
(\ref{xB}), $x_A$ grows faster than $x_B$ for 
$c_1>c_2$. Then, there are the three qualitatively 
different cases: (A) $c_1>c_2>1$, (B) $c_2>c_1>1$, and 
(C) $c_1=c_2>1$. 
In the following, we study each case separately.

\subsection{\texorpdfstring{$c_1>c_2>1$}{c1>c2>1}}

In this case, Eq.~(\ref{dal3}) is dominated by the term arising 
from the 1-form, so we can ignore the 2-form contribution to 
Eq.~(\ref{dal1}). {}From Eq.~(\ref{xA}), we obtain 
\be
e^{-4\alpha-4\sigma-4c_1 \int \frac{\Ve}{M_{\rm pl}^2 \Vep}\mr{d}\phi}
=\frac{(c_1-1) \epsilon_V \Ve}{c_1^2 p_A^2}\,.
\label{1formre}
\ee
Then, the 1-form energy density reduces to 
\be
\rho_A=\frac{c_1-1}{2c_1^2}
\epsilon_V \Ve\,,
\label{rhoA}
\ee
which is nearly constant during inflation. 
{}From Eq.~(\ref{eq:be3}), the ratio between the shear 
and the Hubble expansion rate approaches the value 
\be
\frac{\Sigma}{H} \simeq
\frac{2\rho_A}{3\Ve}=\frac{c_1-1}{3c_1^2} \epsilon_V\,,
\label{SigHA}
\ee
where we used the slow-roll approximation 
$3M_{\rm pl}^2 H^2 \simeq \Ve$.
Substituting Eq.~(\ref{1formre}) into Eq.~(\ref{dal1}), we find
\be
\frac{\mr{d} \phi}{\mr{d} \alpha}=-\frac{M_{\rm pl}^2 \Vep}{\Ve} 
\frac{1}{c_1}\,.
\label{dphial}
\ee
For $c_1>1$, the evolution of $\phi$ slows down 
by the coupling with the 1-form field. 
{}From Eqs.~(\ref{eq:be1}) and (\ref{eq:be2}), the slow-roll 
parameter $\epsilon=-\dot{H}/H^2$ can be estimated as
\be
\epsilon \simeq \frac{\dot{\phi}^2}{2M_{\rm pl}^2 H^2}
+\frac{2\rho_A}{3M_{\rm pl}^2 H^2}\,.
\ee
Since $\dot{\phi}^2/(2M_{\rm pl}^2 H^2)=\epsilon_V/c_1^2$ and 
$2\rho_A/(3M_{\rm pl}^2 H^2)=(c_1-1)\epsilon_V/c_1^2$
from Eqs.~(\ref{rhoA}) and (\ref{dphial}), we have
\be
\epsilon \simeq \frac{\epsilon_V}{c_1}\,,
\ee
and hence $\epsilon<\epsilon_V$ for $c_1>1$.
Then, Eq.~(\ref{SigHA}) is expressed as 
\be
\frac{\Sigma}{H} \simeq \frac{c_1-1}{3c_1} \epsilon\,,
\label{SigH1}
\ee
which is positive and approximately constant  during inflation.
This is the region in which the anisotropic hair is generated 
by the coupling between $\phi$ 
and the 1-form \cite{Watanabe}.

\subsection{\texorpdfstring{$c_2>c_1>1$}{c2>c1>1}}

If $c_2>c_1>1$, then the 2-form field gives the dominant 
contribution to Eq.~(\ref{dal1}). 
Since the solution (\ref{xB}) translates to 
\be
e^{-2\alpha+4\sigma-2c_2 \int \frac{\Ve}{M_{\rm pl}^2 \Vep}\mr{d}\phi}
=\frac{2(c_2-1) \epsilon_V \Ve}{c_2^2 p_B^2}\,,
\label{2formre}
\ee
the 2-form energy density is given by 
\be
\rho_B=\frac{c_2-1}{c_2^2}
\epsilon_V \Ve\,.
\label{rhoB}
\ee
The shear divided by the Hubble expansion rate can be 
estimated as
\be
\frac{\Sigma}{H} \simeq
-\frac{2\rho_B}{3\Ve}=-\frac{2(c_2-1)}{3c_2^2} \epsilon_V\,.
\label{SigHB}
\ee
Applying Eq.~(\ref{2formre}) to Eq.~(\ref{dal1}), we find
\be
\frac{\mr{d} \phi}{\mr{d} \alpha}=-\frac{M_{\rm pl}^2 \Vep}{\Ve} 
\frac{1}{c_2}\,,
\label{dphia2}
\ee
so that the coupling between $\phi$ and the 2-form 
leads to the decrease of inflaton velocity for $c_2>1$.
On using Eqs.~(\ref{rhoB}) and (\ref{dphia2}), the slow-roll 
parameter $\epsilon$ yields
\be
\epsilon \simeq \frac{\dot{\phi}^2}{2M_{\rm pl}^2 H^2}
+\frac{\rho_B}{3M_{\rm pl}^2 H^2}
=\frac{\epsilon_V}{c_2}\,.
\ee
Then, from Eq.~(\ref{SigHB}), it follows that 
\be
\frac{\Sigma}{H} \simeq -\frac{2(c_2-1)}{3c_2}\epsilon\,,
\label{SigHB2}
\ee
which is approximately a negative constant during inflation. 
This result matches with that derived in Ref.~\cite{Ohashi1} 
for the system in which the inflaton is coupled to 
the 2-form field alone.

\subsection{\texorpdfstring{$c_1=c_2>1$}{c1=c2>1}}
\label{c12eq}

If $c_1=c_2>1$, then the differential Eq.~(\ref{dal3}) reduces to
\be
\frac{\mr{d}x}{\mr{d}\phi}=\frac{c_1^2}{\epsilon_V \Ve} \left[ 
2P_A^2  e^{4(c_1-1)\alpha}x^{-1} 
+P_B^2 e^{2(c_1-1) \alpha} 
\right] \frac{\mr{d} \alpha}{\mr{d}\phi}\,,
\label{xphi}
\ee
where $P_A=p_A e^{-2\sigma}$ and 
$P_B=p_Be^{2\sigma}$. 
When we integrate Eq.~(\ref{xphi}) with respect to $\alpha$, 
we deal with the term $\epsilon_V \Ve$ and 
the $\sigma$-dependent terms as constants.
Then, Eq.~(\ref{xphi}) admits the asymptotic 
solution of the form:
\be
x={\cal C}e^{2(c_1-1)\alpha}\,,
\label{xso}
\ee
where the constant ${\cal C}$ obeys
\be
\frac{c_1}{\epsilon_V \Ve} \left( 
\frac{P_A^2}{{\cal C}^2}
+\frac{P_B^2}{2{\cal C}} \right)
=\frac{c_1-1}{c_1}\,.
\label{Cre}
\ee
The solution (\ref{xso}) translates to 
\be
e^{2\alpha+2c_1 \int \frac{\Ve}{M_{\rm pl}^2 \Vep}\mr{d}\phi}
={\cal C}\,.
\label{xso2}
\ee
On using this relation and ignoring the $\sigma$-dependent 
terms in Eq.~(\ref{rhoAB}), the energy densities 
arising from 1- and 2-form fields are given, 
respectively, by 
\be
\rho_A \simeq \frac{P_A^2}{2{\cal C}^2}\,,\qquad 
\rho_B \simeq \frac{P_B^2}{2{\cal C}}\,.
\ee
{}From Eq.~(\ref{Cre}), there is the particular relation 
\be
2\rho_A+\rho_B=\frac{c_1-1}{c_1^2} \epsilon_V \Ve\,.
\label{rhos}
\ee

Substituting Eq.~(\ref{xso2}) into Eq.~(\ref{dal1}) and using Eq.~(\ref{Cre}), 
we obtain
\be
\frac{\mr{d} \phi}{\mr{d} \alpha}=
-\frac{M_{\rm pl}^2 \Vep}{\Ve} 
\frac{1}{c_1}\,,
\label{dphia3}
\ee
and hence $\dot{\phi}^2/(2M_{\rm pl}^2 H^2)=\epsilon_V/c_1^2$.
{}From Eqs.~(\ref{eq:be1}) and (\ref{eq:be2}), the slow-roll parameter 
$\epsilon$ yields
\be
\epsilon \simeq \frac{\dot{\phi}^2}{2M_{\rm pl}^2 H^2}
+\frac{2\rho_A+\rho_B}{3M_{\rm pl}^2H^2}
=\frac{\epsilon_V}{c_1}\,,
\label{epre}
\ee
where we employed the relation (\ref{rhos}) and the slow-roll 
approximation $3M_{\rm pl}^2 H^2 \simeq \Ve$.\\
We define the ratio between 1- and 2-form energy 
densities, as 
\be
r_{AB} \equiv \frac{\rho_A}{\rho_B}
=\frac{P_A^2}{{\cal C}P_B^2}\,.
\label{rABdef}
\ee
Solving Eq.~(\ref{Cre}) for ${\cal C}$ and using Eq.~(\ref{epre}), 
we can write $r_{AB}$ in the form, 
\be
r_{AB}=\frac{4(c_1-1) \epsilon \Ve P_A^2}{c_1 P_B^4} 
\left[ 1+\sqrt{1+\frac{16(c_1-1) \epsilon \Ve P_A^2}{c_1 P_B^4}} \right]^{-1}\,.
\ee
{}From Eqs.~(\ref{rhos}) and (\ref{rABdef}), the 1- and 2-form 
energy densities are expressed, respectively, as
\be
\rho_A=\frac{r_{AB}}{2r_{AB}+1} 
\frac{c_1-1}{c_1} \epsilon \Ve\,,\qquad 
\rho_B=\frac{1}{2r_{AB}+1} 
\frac{c_1-1}{c_1} \epsilon \Ve\,.
\label{rhoABes}
\ee
Then, from Eq.~(\ref{eq:be3}), the ratio between $\Sigma$ 
and $H$ can be estimated as 
\be
\frac{\Sigma}{H} \simeq \frac{2(r_{AB}-1)}{2r_{AB}+1}
\frac{c_1-1}{3c_1}\epsilon \,.
\label{Sigra2}
\ee
In the limit $r_{AB} \to \infty$, we have 
$\rho_A \to (c_1-1) \epsilon \Ve/(2c_1)$ and 
$\rho_B \to 0$ from Eq.~(\ref{rhoABes}).
In this case, the formula (\ref{Sigra2}) reduces to 
$\Sigma/H=(c_1-1)\epsilon/(3c_1)$, 
which is identical to Eq.~(\ref{SigH1}).  
In the limit $r_{AB} \to 0$, we have 
$\rho_A \to 0$, $\rho_B \to (c_1-1) \epsilon \Ve/c_1$, and 
$\Sigma/H=-2(c_1-1)\epsilon/(3c_1)$. 
This value of $\Sigma/H$ is equivalent to 
Eq.~(\ref{SigHB2}) with $c_1=c_2$. 
For $r_{AB}=1$, i.e., $\rho_A=\rho_B$, the anisotropic shear 
vanishes by the compensation of 
$\rho_A$ and $\rho_B$ in Eq.~(\ref{eq:be3}).
In other cases, the anisotropic hair survives during inflation 
with $\Sigma/H$ given by Eq.~(\ref{Sigra2}).

\subsection{Numerical solutions}

To confirm the accuracy of analytic solutions derived above, we perform 
numerical integrations for the quadratic potential 
\be
V_{\rm eff}(\phi)=\frac{1}{2} \mu^2 \phi^2\,,
\label{poten}
\ee
where $\mu$ is a constant having a dimension of mass. 
We introduce the following dimensionless quantities:
\ba 
& &
\hat{v}_A=\frac{v_A}{M_{\rm pl}}\,,\quad
\hat{v}_B=\frac{v_B}{M_{\rm pl}}\,,\quad
\hat{\rho}_A=\frac{\rho_A}{\mu^2 M_{\rm pl}^2}\,,\quad
\hat{\rho}_B=\frac{\rho_B}{\mu^2 M_{\rm pl}^2}\,,\quad
\hat{\phi}=\frac{\phi}{M_{\rm pl}}\,,\quad \nonumber \\
& &
\hat{H}=\frac{\dot{\alpha}}{\mu}=\alpha'\,,\quad 
\hat{m}_v=\frac{m_v}{\mu}\,,
\ea
where a prime represents a derivative with respect to the dimensionless time $\hat{t}=\mu t$. 

For the couplings given by Eq.~(\ref{f12}), 
we can express Eqs.~(\ref{eq:be1})-(\ref{eq:be6}) 
in the forms:
\ba
& & 
\hat{H}=\sqrt{\sigma'^2+\frac{1}{6}\hat{\phi}'^2
+\frac{1}{6}\hat{\phi}^2+\frac{1}{3} \hat{\rho}_A
+\frac{1}{3} \hat{\rho}_B}\,,\label{Heq}\\
& & 
\hat{H}'=-3\sigma'^2-\frac{1}{2} \phi'^2
-\frac{2}{3} \hat{\rho}_A
-\frac{1}{3} \hat{\rho}_B\,,\label{dHeq} \\
& &
\sigma''=-3 \hat{H} \sigma'+\frac{2}{3} \hat{\rho}_A
-\frac{2}{3} \hat{\rho}_B\,,\\
& &
\hat{\phi}''=-3\hat{H} \hat{\phi}'-\hat{\phi}
+2c_1 \hat{\phi} \hat{\rho}_A+c_2 \hat{\phi} \hat{\rho}_B\,.
\label{dphieq}
\ea
{}From Eqs.~(\ref{drhoA}) and (\ref{drhoB}), the 
1- and 2-form energy densities obey the differential equations:
\ba
\hat{\rho}_A' &=& -4\hat{\rho}_A \left( \hat{H}
+\sigma'+\frac{c_1}{2} \hat{\phi} \hat{\phi}' 
\right)-2\hat{m}_v \sqrt{\frac{\hat{\rho}_A \hat{\rho}_B}
{f_1 f_2}}\,,\label{hatrhoA}\\
\hat{\rho}_B' &=& -2\hat{\rho}_B \left( \hat{H}
-2\sigma'+\frac{c_2}{2} \hat{\phi} \hat{\phi}' 
\right)+2\hat{m}_v \sqrt{\frac{\hat{\rho}_A \hat{\rho}_B}
{f_1 f_2}}\,.
\label{hatrhoB}
\ea
For $\hat{m}_v=0$ the last terms on the right hand sides of 
Eqs.~(\ref{hatrhoA}) and (\ref{hatrhoB}) vanish, so
the anisotropic inflationary dynamics is known 
by integrating Eqs.~(\ref{dHeq})-(\ref{hatrhoB}) 
with Eq.~(\ref{Heq}) for given initial values of 
$\hat{\phi}', \hat{\phi}, \sigma, \sigma', \hat{\rho}_A, 
\hat{\rho}_B$.

\begin{figure}[tb]
\begin{center}
\includegraphics[height=2.9in,width=3.0in]{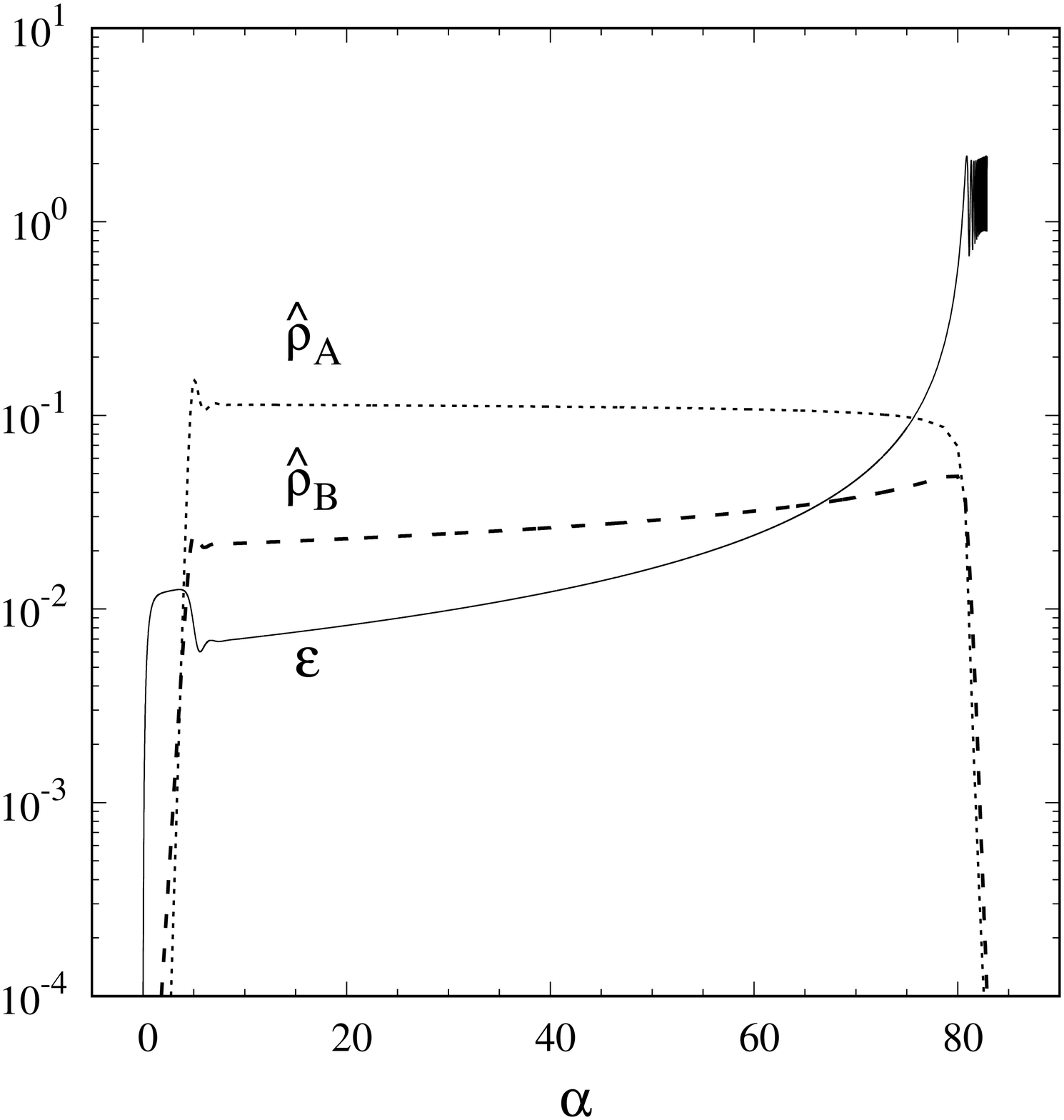}
\includegraphics[height=2.9in,width=3.0in]{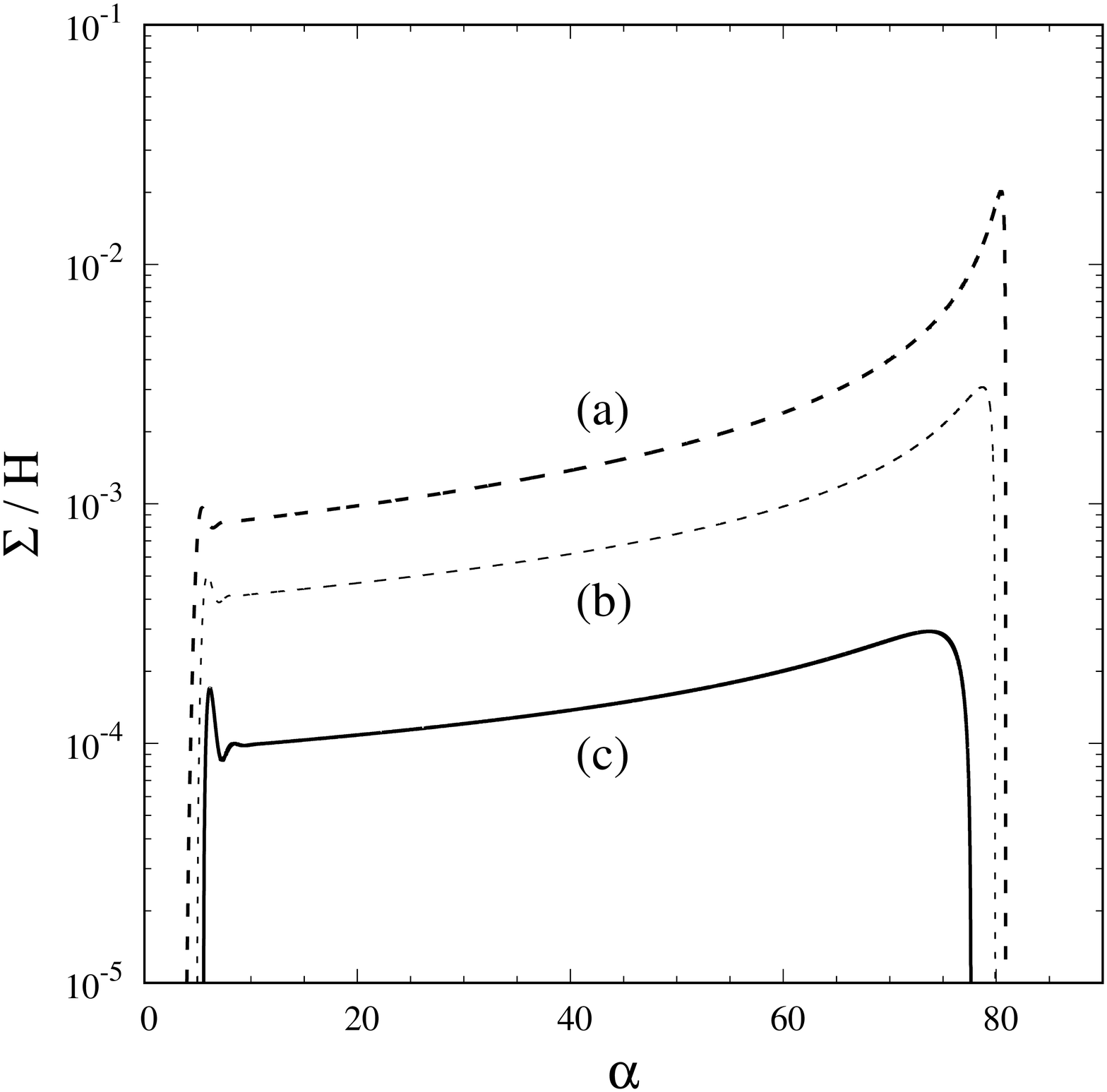}
\end{center}
\caption{\label{fig1}
(Left) Evolution of $\epsilon$, $\hat{\rho}_A$ and $\hat{\rho}_B$ 
versus the number of e-foldings $\alpha=\ln a$ for the potential 
(\ref{poten}) with $m_v=0$ and $c_1=c_2=2$. 
The initial conditions are chosen to be $\hat{\phi}'=0$, $\hat{\phi}=13$, 
$\sigma=0$, $\sigma'=10^{-10}$, $\hat{\rho}_A=2.5 \times 10^{-8}$, 
and $\hat{\rho}_B=10^{-5}$ at the onset of integration ($\alpha=0$).
(Right) The case (a) corresponds to the evolution of $\Sigma/H$ versus 
$\alpha$ for the same model parameters and initial conditions 
as those in the left panel. 
In cases (b) and (c), the difference from case (a) is only the initial condition 
of $\hat{\rho}_A$, such that 
(b) $\hat{\rho}_A=3.5 \times 10^{-9}$ and 
(c)  $\hat{\rho}_A=1.5 \times 10^{-9}$ at $\alpha=0$.}
\end{figure}

In the left panel of Fig.~\ref{fig1}, we exemplify the evolution of 
$\epsilon$, $\hat{\rho}_A$, and $\hat{\rho}_B$ for $m_v=0$ and $c_1=c_2=2$.
After the initial transient period, the Universe enters the stage in 
which both $\hat{\rho}_A$ and $\hat{\rho}_B$ are nearly 
constant. In this case, the ratio (\ref{rABdef}) is 
$r_{AB}=\hat{\rho}_A/\hat{\rho}_B=4.89$ at $\alpha=20$. 
Inflation ends around $\alpha=80.4$ at which the slow-roll 
parameter $\epsilon$ exceeds 1. 
The 1- and 2-form energy densities start to decrease 
around the end of inflation. 
The case (a) in the right panel of Fig.~\ref{fig1} corresponds 
to the same model parameters and initial conditions as those 
in the left panel. We observe that the anisotropic shear
survives during inflation with the slow increase of $\Sigma/H$. 
On using the values $r_{AB}=4.89$ and $\epsilon=8.22 \times 10^{-3}$ 
at $\alpha=20$, the analytic formula (\ref{Sigra2}) gives 
$\Sigma/H \simeq 0.120 \epsilon=9.89 \times 10^{-4}$, 
which exhibits good agreement with its numerical value.

The case (b) in the right panel of Fig.~\ref{fig1} corresponds 
to $r_{AB}=1.77$ and $\epsilon=8.27 \times 10^{-3}$ at $\alpha=20$, 
so that $\Sigma/H \simeq 5.65 \times 10^{-2} \epsilon=4.68 \times 10^{-4}$ from Eq.~(\ref{Sigra2}).  
In this case, anisotropic inflation occurs with the smaller 
ratio $\Sigma/H$ compared to case (a) by reflecting the fact 
that $r_{AB}$ is smaller.
In case (c), the quantity $r_{AB}$ at $\alpha=20$ 
is  $r_{AB}=1.13$, so this is quite close to the border value 
$r_{AB}=1$ at which $\Sigma/H$ changes its sign.
If the value of $r_{AB}$ is smaller than 1 before 
the solutions reach the regime with the nearly constant 
ratio $\Sigma/H$, we find that $\Sigma/H$ is negative 
during anisotropic inflation.

For $r_{AB}$ close to 0, our numerical simulations show that 
there exists the anisotropic inflationary period with the negative 
value $\Sigma/H \simeq -2(c_1-1)\epsilon/(3c_1)$. 
The accuracy of the analytic formula (\ref{Sigra2}) 
is also numerically checked for arbitrary positive values of $r_{AB}$.
Thus, for $c_1=c_2>1$, we have confirmed that 
anisotropic inflation with the shear in the range 
\be
-\frac{2(c_1-1)}{3c_1}\epsilon<\frac{\Sigma}{H}
<\frac{c_1-1}{3c_1}\epsilon\,,
\ee
is indeed realized. 
There is a specific case around $r_{AB}=1$ in which 
$\Sigma/H$ is vanishingly small due to the compensation 
of 1- and 2-form contributions to the shear.

In Fig.~\ref{fig1}, we find that $\Sigma/H$ 
starts to decrease around the end of inflation. 
After inflation, the Universe enters the reheating stage 
in which the inflaton field oscillates around the 
potential minimum.
The precise evolution of $\Sigma$ during reheating 
depends on how the scalar and form fields decay to 
radiation, but as long as 
$\rho_A$ and $\rho_B$ quickly decrease toward 0, 
Eq.~(\ref{eq:be3}) shows that $\Sigma$ decreases 
in proportion to $a^{-3}$.

The numerical results of Fig.~\ref{fig1} correspond to the 
case $c_1=c_2>1$, but we also confirmed that, 
for $c_1>c_2>1$ and $c_2>c_1>1$, there are 
the anisotropic inflationary periods in which $\Sigma/H$ 
is given by Eqs.~(\ref{SigH1}) and (\ref{SigHB2}) respectively. 
These properties hold for other choices 
of the inflaton potential $V_{\rm eff}(\phi)$ as well.

\section{Anisotropic inflation for coupled 1-form and 2-form fields}
\label{coupledbf}

In this section, we study whether anisotropic inflation in which 
both $\rho_A$ and $\rho_B$ are nearly constant can 
be realized by the presence of the non-vanishing coupling 
$m_v$ between 1- and 2-form fields. 

For $m_v=0$, we showed that this is possible for the 
functions (\ref{f12}) with $c_1=c_2>1$. 
In the regime where the anisotropic hair is present, 
there is the particular relation (\ref{xso2}) and hence 
$\mr{d}\alpha/\mr{d}\phi=-c_1 \Ve/(M_{\rm pl}^2 \Vep)$. 
Then, after ignoring the $\dot{\sigma}$ term, the first terms 
on the right hand sides of Eqs.~(\ref{drhoA}) and (\ref{drhoB}) vanish. 
To estimate the effect of non-vanishing coupling $m_v$ 
on the solutions derived for $m_v=0$ and $c_1=c_2>1$, 
we write 
$\rho_A$ and $\rho_B$ in the forms 
$\rho_A=\bar{\rho}_A+\delta \rho_A(t)$ and 
$\rho_B=\bar{\rho}_B+\delta \rho_B(t)$, where 
$\bar{\rho}_A$ and $\bar{\rho}_B$ are constants.
On using the relation (\ref{xso2}) with $c_1=c_2$, the 
couplings given by Eq.~(\ref{f12}) evolve as 
$f_1=\bar{f}_1 a^{-4}$ and $f_2=\bar{f}_2 a^{-2}$, where 
$\bar{f}_1$ and $\bar{f}_2$ are constants. 
Then, the homogeneous perturbations $\delta \rho_A(t)$ and 
$\delta \rho_B(t)$ obey 
\be
\dot{\delta \rho_A}=-2m_v \sqrt{\frac{\bar{\rho}_A \bar{\rho}_B}
{\bar{f}_1 \bar{f}_2}}\,a^3\,,
\qquad
\dot{\delta \rho_B}=2m_v \sqrt{\frac{\bar{\rho}_A \bar{\rho}_B}
{\bar{f}_1 \bar{f}_2}}\,a^3\,.
\label{deltarhoAB}
\ee
Approximating the inflationary background as the de-Sitter 
expansion, i.e., $a=e^{Ht}$, the integrated solutions to 
Eq.~(\ref{deltarhoAB}) are 
\be
\delta \rho_A=-\frac{2m_v}{3H}
 \sqrt{\frac{\bar{\rho}_A \bar{\rho}_B}
{\bar{f}_1 \bar{f}_2}}\,e^{3Ht}\,,\qquad 
\delta \rho_B=\frac{2m_v}{3H}
 \sqrt{\frac{\bar{\rho}_A \bar{\rho}_B}
{\bar{f}_1 \bar{f}_2}}\,e^{3Ht}\,,
\label{drhoABes}
\ee
where we dropped the integration constants. {}From Eq.~(\ref{drhoABes}), 
the non-vanishing coupling $m_v$ 
leads to the deviation from the solutions 
$\rho_A=\bar{\rho}_A$ 
and $\rho_B=\bar{\rho}_B$. 
In other words, the anisotropic inflationary solutions supported 
by uncoupled 1- and 2-form fields for $c_1=c_2>1$ 
tend to disappear by taking into account their interactions.

Let us discuss the other choices of couplings $f_1$ and $f_2$ 
for the realization of solutions $\dot{\rho}_A \simeq 0$ and 
$\dot{\rho}_B \simeq 0$. For $c_1$ and $c_2$ of order unity, the 
terms $\dot{\alpha}+\dot{\sigma}+\dot{f}_1/(4f_1)$ and 
$\dot{\alpha}-2\dot{\sigma}+\dot{f}_2/(2f_2)$ in Eqs.~(\ref{drhoA}) 
and (\ref{drhoB}) are at most of order $H$.
Then, the conditions $\dot{\rho}_A \simeq 0$ and 
$\dot{\rho}_B \simeq 0$ translate to  
$\rho_A/\rho_B \propto m_v^2/(H^2 f_1 f_2)$ and 
$\rho_B/\rho_A \propto m_v^2/(H^2 f_1 f_2)$, respectively.  
During inflation, the compatibility of these two conditions 
implies that $f_1 f_2={\rm constant}$, i.e., 
\be
c_2=-2c_1\,.
\label{c12}
\ee
In this case, the mass term defined by 
\be
\bar{m}_v=\frac{m_v}{\sqrt{f_1 f_2}}\,,
\ee
is constant.

\subsection{Analytic solutions}
\label{analysec}

We derive analytic solutions to the shear $\Sigma=\dot{\sigma}$ 
during inflation for the constants $c_1$ and $c_2$ 
satisfying the condition (\ref{c12}). 
Then, we can express Eqs.~(\ref{drhoA}) and (\ref{drhoB}) 
in the following forms:
\ba
\dot{\rho}_A
&=&-4 \rho_A \left( \dot{\alpha}+\dot{\sigma}
+c_1 \frac{\Ve}{M_{\rm pl}^2 \Vep}\dot{\phi} \right)
-2\bar{m}_v \sqrt{\rho_A \rho_B}\,,
\label{trhoAd}\\
\dot{\rho}_B
&=&-2\rho_B \left( \dot{\alpha}-2\dot{\sigma}
-2c_1 \frac{\Ve}{M_{\rm pl}^2 \Vep}\dot{\phi} \right)
+2\bar{m}_v \sqrt{\rho_A \rho_B}\,.
\label{trhoBd}
\ea
Imposing the conditions $\dot{\rho}_A=0$ and 
$\dot{\rho}_B=0$, 
we obtain the following relation 
\be
\frac{\rho_B}{\rho_A}=
\frac{9H^2}{4\bar{m}_v^2} \left[ 
\sqrt{1+\frac{4\bar{m}_v^2}{9H^2}}-1 \right]^2\,.
\label{rhoBA}
\ee
If $\bar{m}_v^2/H^2 \ll 1$, then Eq.~(\ref{rhoBA}) 
reduces to 
\be
\frac{\rho_B}{\rho_A} \simeq \frac{\bar{m}_v^2}
{9H^2} \ll 1\,.
\label{rhora}
\ee
In another limit $\bar{m}_v^2/H^2 \gg 1$, the ratio 
$\rho_B/\rho_A$ approaches 1. 
In the latter regime, we can set $\rho_B \simeq \rho_A$ in  
Eq.~(\ref{trhoAd}), so that 
\be
\dot{\rho}_A \simeq 
-2\rho_A  \left( 2H+2\dot{\sigma}
+2c_1 \frac{\Ve}{M_{\rm pl}^2 \Vep}\dot{\phi} 
+\bar{m}_v \right)\,.
\ee
Since $\bar{m}_v$ is much larger than $H$, 
the exponential decrease of $\rho_A$ occurs during inflation. 
Hence the anisotropic shear does not survive
in the regime $\bar{m}_v^2/H^2 \gg 1$.  
As we will see below, this is not the case for $\bar{m}_v^2$ 
smaller than the order of $H^2$.

Let us explore the dynamics of anisotropic inflation in the 
regime $\bar{m}_v^2/H^2 \ll 1$. 
On using Eq.~(\ref{rhora}), Eqs.~(\ref{trhoAd}) and 
(\ref{trhoBd}) reduce, respectively, to 
\ba
\dot{\rho}_A &\simeq& -4\rho_A \left( \dot{\alpha}+\dot{\sigma}
+c_1 \frac{\Ve}{M_{\rm pl}^2 \Vep}\dot{\phi}
+\frac{\bar{m}_v^2}{6H} \right)\,,\label{rhoAap1} \\
\dot{\rho}_B &\simeq&  4\rho_B \left( \dot{\alpha}+\dot{\sigma}
+c_1 \frac{\Ve}{M_{\rm pl}^2 \Vep}\dot{\phi} \right)\,.
\label{rhoBap1} 
\ea
{}From Eq.~(\ref{rhoBap1}), the condition 
$\dot{\rho}_B=0$ holds for 
\be
\dot{\alpha}+\dot{\sigma}
+c_1 \frac{\Ve}{M_{\rm pl}^2 \Vep}\dot{\phi} 
=0\,.
\label{rhore} 
\ee
Substituting Eq.~(\ref{rhore}) into Eq.~(\ref{rhoAap1}), 
it follows that 
\be
\dot{\rho}_A = -\frac{2\bar{m}_v^2}{3H} \rho_A\,,
\ee
whose solution is 
\be
\rho_A =\rho_{A0} \exp \left( -\int_0^{\alpha} 
\frac{2\bar{m}_v^2}{3H^2}\,\mr{d} \tilde{\alpha} \right)\,,
\label{rhoex}
\ee
where $\rho_{A0}$ is an integration constant. 
The critical time $t_c$ after which $\rho_A$ is 
subject to the exponential suppression is identified 
by the moment at which the integral 
$\int_0^{\alpha} 2\bar{m}_v^2/(3H^2)\,{\rm d}\tilde{\alpha}$ 
reaches the order 1. 
For $t<t_c$, $\rho_A$ stays nearly constant with 
$\rho_B \simeq \rho_A\,\bar{m}_v^2/(9H^2)$.
Then, from Eq.~(\ref{eq:be3}), it is possible to realize 
anisotropic inflation characterized by 
\be
\frac{\Sigma}{H} \simeq 
\frac{2\rho_A}{3V_{\rm eff}} \left( 
1-\frac{\bar{m}_v^2}{9H^2} \right)\,,
\label{Siges}
\ee
where we employed the slow-roll approximation 
$3M_{\rm pl}^2 H^2 \simeq V_{\rm eff}$ 
together with the condition $\rho_A \ll \Ve$.
Using the slow-roll approximation in
Eq.~(\ref{eq:be6}), we obtain 
\be
\frac{\mr{d}\phi}{\mr{d}\alpha} \simeq -\frac{M_{\rm pl}^2 \Vep}{\Ve}
+\frac{4c_1 \rho_A}{\Vep} 
\left( 1-\frac{\bar{m}_v^2}{9H^2} \right)\,.
\label{dphial2}
\ee
{}From Eqs.~(\ref{rhore}), (\ref{Siges}), and (\ref{dphial2}), 
the 1-form energy density can be expressed as 
\be
\rho_A \simeq \frac{c_1-1}{2c_1^2} \epsilon_V \Ve
\left( 1-\frac{\bar{m}_v^2}{9H^2} \right)^{-1}\,,
\label{rhoA2}
\ee
where we ignored the slow-roll corrections to $\rho_A$ 
higher than the linear order in $\epsilon_V$. 
The positivity of $\rho_A$ requires that $c_1>1$.
In the limit $\bar{m}_v^2 \to 0$, the solution (\ref{rhoA2})
recovers Eq.~(\ref{rhoA}) derived for $\bar{m}_v=0$ 
and $c_1>c_2>1$. 
Substituting Eq.~(\ref{rhoA2}) into Eq.~(\ref{dphial2}), 
we have 
\be
\frac{\mr{d} \phi}{\mr{d} \alpha} \simeq 
-\frac{M_{\rm pl}^2 \Vep}{\Ve} 
\frac{1}{c_1}\,,
\label{phialf}
\ee
which is analogous to Eq.~(\ref{dphial}). 
{}From Eqs.~(\ref{eq:be1}) and (\ref{eq:be2}), the 
slow-roll parameter $\epsilon=-\dot{H}/H^2$ 
is expressed as
\be
\epsilon \simeq \frac{\epsilon_V}{c_1} \left( 
1+\frac{c_1-1}{6c_1} \frac{\bar{m}_v^2}{H^2} 
\right)\,,
\label{epf}
\ee
where we exploited Eq.~(\ref{phialf}) and picked up 
the leading-order term in the expansion of $\bar{m}_v^2/H^2$. 
Applying Eqs.~(\ref{rhoA2}) and (\ref{epf}) to  
Eq.~(\ref{Siges}), we obtain
\be
\frac{\Sigma}{H}  \simeq \frac{c_1-1}{3c_1^2} 
\epsilon_V \simeq 
\frac{c_1-1}{3c_1} \epsilon 
\left( 1-\frac{c_1-1}{6c_1} \frac{\bar{m}_v^2}
{H^2}\right)\,.
\label{SigHg}
\ee
Hence the anisotropic shear can survive during 
inflation for the coupled system of 1- and 2-forms. 
This period ends after the 
exponential decrease of $\rho_A$ characterized by 
Eq.~(\ref{rhoex}) becomes significant at $t>t_c$.
For $c_1>1$, the 2-form energy density provides 
the negative contribution to Eq.~(\ref{SigHg}), 
but the ratio $\Sigma/H$ remains positive during anisotropic 
inflation due to the first approximate equality of Eq.~(\ref{SigHg}).

\begin{figure}[h]
\begin{center}
\includegraphics[height=2.9in,width=3.0in]{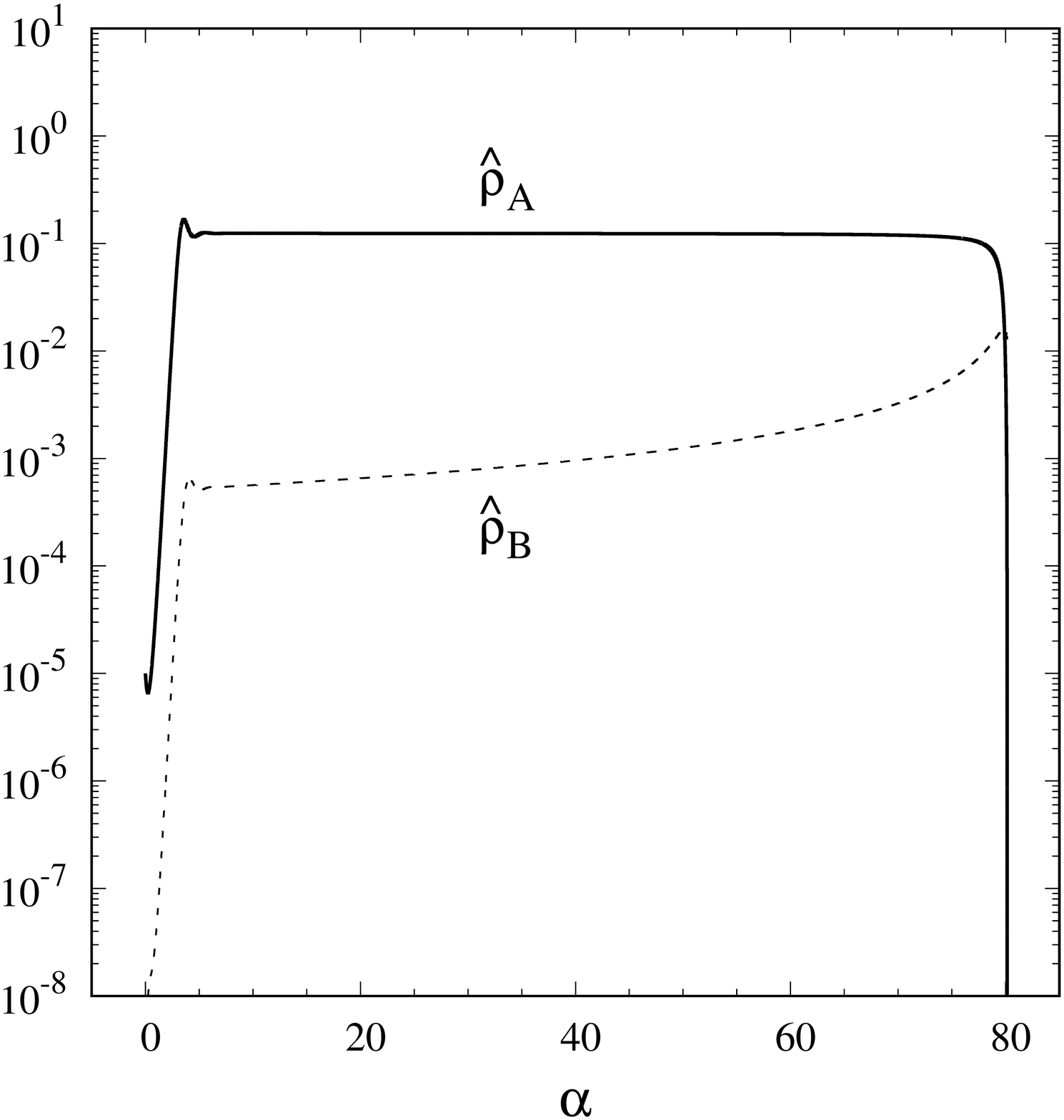}
\includegraphics[height=2.9in,width=3.0in]{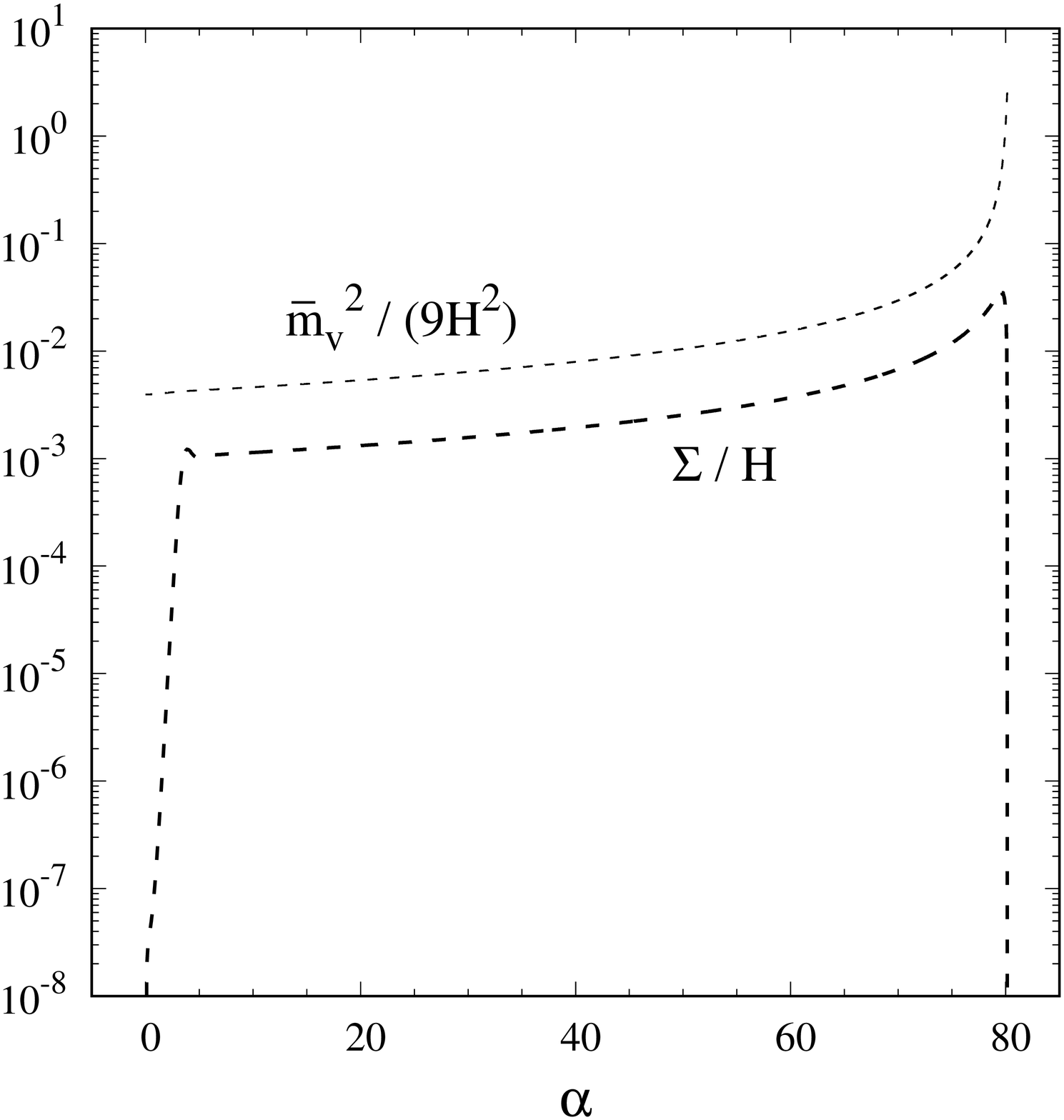}
\end{center}
\caption{\label{fig2}
Evolution of $\hat{\rho}_A=\rho_A/(\mu^2 M_{\rm pl}^2), 
\hat{\rho}_B=\rho_B/(\mu^2 M_{\rm pl}^2)$ (left) and 
$\bar{m}_v^2/(9H^2), \Sigma/H$ (right) 
versus $\alpha=\ln a$ for 
the potential (\ref{poten}) with 
$\bar{m}_v=\mu$, $c_1=2$, and $c_2=-4$. 
The initial conditions are chosen to be 
$\hat{\phi}'=0$, $\hat{\phi}=13$, 
$\sigma'=10^{-10}$, $\hat{\rho}_A=10^{-5}$, 
and $\hat{\rho}_B=10^{-10}$
at the onset of integration ($\alpha=0$).}
\end{figure}

%
\subsection{Numerical solutions}

To confirm the analytic estimation given in Sec.~\ref{analysec}, 
we consider the quadratic potential (\ref{poten}) and 
numerically integrate Eqs.~(\ref{hatrhoA})-(\ref{hatrhoB}) with 
Eqs.~(\ref{Heq})-(\ref{dphieq}) for $\bar{m}_v \neq 0$.
In the left panel of Fig.~\ref{fig2}, we show one example for the 
evolution of $\hat{\rho}_A$ and $\hat{\rho}_B$ versus 
$\alpha=\ln a$ for $\bar{m}_v=\mu$, $c_1=2$, and $c_2=-4$. 
This is the case in which the condition $\bar{m}_v^2/H^2 \ll 1$ 
is satisfied at the onset of integration.
As estimated by Eq.~(\ref{rhoex}) in the regime 
$\int_0^{\alpha} 2\bar{m}_v^2/(3H^2)\,\mr{d}\tilde{\alpha} \ll 1$, 
there exists the anisotropic inflationary period in which 
$\rho_A$ stays nearly constant. 
Once the integral $\int_0^{\alpha} 2\bar{m}_v^2/(3H^2)\,
{\rm d}\tilde{\alpha}$ reaches the order 1, the 1-form 
energy density $\rho_A$ is subject to the exponential suppression. 
Soon after this decrease of $\rho_A$, the inflationary period 
ends around $\alpha=79.7$ in the numerical simulation of 
Fig.~\ref{fig2}.

In the regime $\bar{m}_v^2 \ll H^2$, the 2-form energy density 
slowly grows as $\rho_B \simeq \rho_A\,\bar{m}_v^2/(9H^2)$ 
with the decrease of $H$.
Eventually, $\rho_B$ catches up with 
$\rho_A$ around the moment at which 
$\bar{m}_v^2/(9H^2)$ exceeds the order 1. 
As estimated by Eq.~(\ref{SigHg}), the right panel of 
Fig.~\ref{fig2} shows that there exists the period of 
anisotropic inflation in which the ratio $\Sigma/H$ 
stays nearly a constant. 
The energy density $\rho_A$ is the dominant source for 
the shear, but $\rho_B$ also 
contributes to $\Sigma$. 
In the numerical simulation of Fig.~\ref{fig2} we chose the 
specific values $c_2=-2c_1=-4$, but for the general 
coupling constants satisfying
\be
c_2=-2c_1 \quad {\rm and} \quad c_1>1\,,
\label{c21}
\ee
we numerically confirmed the existence of anisotropic hairs endowed with 
the coupled 1- and 2-form fields.
As we discussed in Sec.~\ref{analysec}, we require the condition 
$\bar{m}_v^2/H^2 \ll 1$ to avoid the exponential suppression 
of $\rho_A$ and hence
$\rho_B \simeq \rho_A\,\bar{m}_v^2/(9H^2) \ll \rho_A$ 
during most stage of anisotropic inflation.
Since $\rho_B$ cannot dominate over $\rho_A$ to keep 
the condition $\dot{\rho}_A \simeq 0$, the couplings 
satisfying $c_2>1$ and $c_2=-2c_1$ do not sustain the
anisotropic shear. We recall that, for $m_v=0$, the anisotropic hair 
induced by both 1- and 2-form fields survives only for $c_1=c_2>1$. 
The non-vanishing coupling $m_v$ allows the possibility for realizing 
new hairy solutions for the negative constant $c_2$ 
satisfying the condition (\ref{c21}).

\section{Conclusions}
\label{sec:Conclusions}

We studied the dynamics of anisotropic inflation in gauge-invariant 
coupled $p$-form theories with parity invariance. 
The 3-form coupled to the scalar field $\phi$ generates the 
effective scalar potential after integrating out 
interacting Lagrangians from the action. 
As a result, the reduced action of coupled $p$-forms minimally 
coupled to gravity is of the form (\ref{eq:LT}), 
which contains the interacting term $-m_v B \tilde{F}/2$ 
between 1- and 2-forms. 
In Sec.~\ref{sec:background}, we derived the equations of motion on 
the anisotropic background (\ref{anisotropic-metric}) to study the 
evolution of the cosmic shear during inflation.

If $m_v=0$,  it is known that the couplings $-f_1(\phi)F^2/4$ and 
$-f_2(\phi)H^2/12$ in the action (\ref{eq:LT}) can separately sustain 
the anisotropic shear during slow-roll inflation 
for the couplings (\ref{f12}) \cite{Watanabe,Ohashi1}.
When these two couplings coexist, the presence
of an anisotropic inflationary attractor was shown in 
Ref.~\cite{Ito} for the exponential potential 
$\Ve (\phi)=V_0e^{\lambda \phi}$.
Without specifying any inflaton potential, we derived the general 
analytic formulas of the shear $\Sigma$ to the Hubble expansion rate $H$ 
during inflation in the presence of two couplings mentioned above. 
As we showed in Sec.~\ref{unsec}, there are 
three qualitatively different cases depending on the coupling constants 
$c_1$ and $c_2$: (A) $c_1>c_2>1$, (B) $c_2>c_1>1$, and (C) $c_1=c_2>1$. 
The case (C) is particularly of interest, as both 1- and 2-forms contribute 
to the shear. As we see in the formula (\ref{Sigra2}), there is a special case 
in which $\Sigma/H$ vanishes at $r_{AB}=1$ due to the compensation 
of 1- and 2-form contributions to the shear.
In Fig.~\ref{fig1}, we confirmed that our analytic formulas are sufficiently 
accurate during the anisotropic inflationary period driven by the 
quadratic potential (\ref{poten}).

When $m_v \neq 0$, we explored the possibility for realizing anisotropic inflation 
supported by both 1- and 2-forms. 
In Sec.~\ref{coupledbf}, we first showed that the anisotropic shear 
present for the couplings $c_1=c_2>1$ with $m_v=0$ tends to disappear 
by the non-vanishing coupling $m_v$ between 1- and 2-forms. 
However, for the couplings satisfying $c_2=-2c_1$ and $c_1>1$, 
we found a new class of anisotropic inflationary solutions along which 
both $\rho_A$ and $\rho_B$ are approximately constant. 
In the regime $\bar{m}_v^2/H^2 \ll 1$, $\rho_B$ is sustained by $\rho_A$ 
according to the relation (\ref{rhora}).
As we observe in Eq.~(\ref{rhoex}), $\rho_A$ is nearly constant by the time at which 
the integral $\int_0^{\alpha} 2\bar{m}_v^2/(3H^2)\,{\rm d}\tilde{\alpha}$
reaches the order 1. We showed that the ratio $\Sigma/H$ is analytically 
given by Eq.~(\ref{SigHg}) during slow-roll anisotropic inflation. 
Our analytic formulas were also numerically confirmed 
for the inflaton potential (\ref{poten}), see Fig.~\ref{fig2}.

There are several issues we did not address 
in this paper. First, it will be of interest to estimate 
the effect of the inflationary anisotropic shear 
on the primordial power spectra of scalar and tensor perturbations as well as on the non-linear 
estimator $f_{\rm NL}$ of primordial non-Gaussianities with/without the 
interactions between 1- and 2-forms. 
In particular, the anisotropy parameter $g_*$ in the scalar power spectrum 
would be subject to change by the partial compensation 
of 1- and 2-form contributions to the shear. 
They can be exploited to confront the models of 
anisotropic inflation with the observational data of 
CMB temperature anisotropies. 
Finally, the application of our coupled $p$-form 
theories to the late-time cosmology, in particular, 
to the dynamics of dark energy and associated fixed points 
with their stabilities will be also interesting.

\section*{Acknowledgments}
This work was partly supported by COLCIENCIAS grant 110671250405 RC FP44842-103-2016 
and by COLCIENCIAS -- DAAD grant 110278258747. JPBA acknowledge support from 
Universidad Antonio Nari\~no grant  2017239 and thanks Tokyo University of Science for kind hospitality at early stages of this project. 
RK is supported by the Grant-in-Aid for Young Scientists B of the JSPS No.\,17K14297.  ST is supported by the Grant-in-Aid for Scientific Research Fund of the 
JSPS No.~16K05359 and MEXT KAKENHI Grant-in-Aid for Scientific Research on Innovative Areas ``Cosmic Acceleration'' (No.\,15H05890). 



\appendix

\end{document}